\documentclass[twocolumn]{aastex631}
\usepackage{amsmath}
\usepackage{threeparttable}

\received{XXX}
\revised{YYY}
\accepted{ZZZ}
\submitjournal{ApJ}
\shorttitle{Standardizing long GRBs}
\shortauthors{Wang et al.}

\begin{document}

\title{Standardized long gamma-ray bursts as a cosmic distance indicator}

\author[0000-0003-4157-7714]{F. Y. Wang}
\affiliation{School of Astronomy and Space Science, Nanjing University, Nanjing 210023, China}
\affiliation{Key Laboratory of Modern Astronomy and Astrophysics (Nanjing University), Ministry of Education, Nanjing 210023, China}

\author[0000-0002-5819-5002]{J. P. Hu}
\affiliation{School of Astronomy and Space Science, Nanjing University, Nanjing 210023, China}

\author[0000-0001-6545-4802]{G. Q. Zhang}
\affiliation{School of Astronomy and Space Science, Nanjing University, Nanjing 210023, China}

\author[0000-0002-7835-8585]{Z. G. Dai}
\affil{Department of Astronomy, School of Physical Sciences, University of Science and Technology of China, Hefei 230026, Anhui, China}
\affil{School of Astronomy and Space Science, Nanjing University, Nanjing 210023, China}

\correspondingauthor{F. Y. Wang}
\email{fayinwang@nju.edu.cn}

\begin{abstract}
Gamma-ray bursts (GRBs) are the most luminous explosions and can be detectable out to the edge of Universe. It
has long been thought they can extend the Hubble diagram to very high redshifts. Several correlations
between temporal or spectral properties and GRB luminosities have been proposed to make GRBs cosmological tools. However, those correlations cannot be properly standardized. In this paper, we select a long GRB sample with X-ray plateau phases produced by electromagnetic dipole emissions from central new-born magnetars. A tight correlation is found between the plateau luminosity and the end time of the plateau in X-ray afterglows out to the redshift $z=5.91$.
We standardize these long GRBs X-ray light curves to a universal behavior by this correlation for the first time, with a luminosity dispersion of 0.5 dex. The derived distance-redshift relation of GRBs is in agreement with the standard $\Lambda$CDM model both at low and high redshifts. The evidence of accelerating universe from this GRB sample is $3\sigma$, which is the highest statistical significance from GRBs to date.
\end{abstract}

\keywords{Gamma-ray bursts (629)	--- Cosmological parameters (339) --- Magnetars (992)	}


\section{Introduction} \label{sec:intro}

The cosmological-constant ($\Lambda$) cold dark matter (CDM) model successfully describes the majority of cosmological observations\citep{2016A&A...594A..13P,2020A&A...641A...6P}. However, the $\Lambda$CDM model is challenged by $H_0$ tension \citep{2019ApJ...876...85R} and high-redshift probes \citep{2019NatAs...3..272R}. So one urgently needs distance indicators to probe the expansion of universe at high redshifts. Gamma-ray bursts (GRBs) are short and intense pulses of soft gamma rays emitting up to 10$^{54}$ erg energy \citep{2006RPPh...69.2259M,2009ARA&A..47..567G,2015PhR...561....1K}. They cover a very wide redshift range, up to
$z=9.4$, which makes them as appealing cosmological probes, complementary to type Ia supernovae and cosmic microwave
background \citep[for a recent review, see][]{2015NewAR..67....1W}. The bimodal duration distribution leads to a classification
of them into two types, i.e. ``long" bursts with duration $>2$ s and ``short" bursts with duration $<2$ s. The
progenitors of long GRBs are thought to be massive stars, while short GRBs arise from compact object binary mergers \citep{2009ARA&A..47..567G,Abbott2017}.

One outstanding question is how to standardize GRBs as a reliable cosmic distance indicator.
Interestingly, some works have shown that long GRBs can be potentially used to extend the Hubble diagram out to high
redshifts \citep{2001ApJ...562L..55F,2004ApJ...612L.101D,2004ApJ...613L..13G,2004ApJ...616..331G,2007ApJ...660...16S, 2005ApJ...633..611L}.
However, due to the diversity of light curves, a reliable method to standardize them has not yet been established,
though recent work provides encouraging results \citep{2009MNRAS.400..775C,2013ApJ...774..157D,2014ApJ...783..126P,2016A&A...585A..68W,2017A&A...598A.113D,2019MNRAS.486L..46A,2019ApJ...887...13F,2019ApJS..245....1T,2020arXiv201205627X,2021ApJ...908..181M,Khadka21}.

Interestingly, a significant fraction of long GRBs in the Neil Gehrels Swift Observatory sample has a plateau
phase in the X-ray light curves \citep{2006ApJ...642..354Z,2006ApJ...642..389N}. This phase is usually believed to be continuous energy injection from a newly born
rapidly spinning,
strongly magnetized neutron star called ``millisecond magnetar" \citep{1998A&A...333L..87D,2001ApJ...552L..35Z,2011MNRAS.413.2031M}. The energy reservoir of a newly born magnetar is
rotational
energy, which is given by
\begin{equation}
E_{\rm rot} = \frac{1}{2} I \Omega_{0}^{2}
\simeq 2 \times 10^{52}~{\rm erg}~
M_{1.4} R_6^2 P_{0,ms}^{-2},
\label{Erot}
\end{equation}
where $I$ is the stellar moment of inertia, $\Omega_0 = 2\pi/P_0$ is
the initial angular frequency of the magnetar with period $P_0$ in units of millisecond, $R_6=R/10^6$ cm is the typical radius of
magnetar in units of $10^6$ cm and $M_{1.4} =
M/1.4M_\odot$ is the magnetar mass. The rotational energy is released as gravitational wave and
electromagnetic radiation,
causing the magnetar to spin down.
Assuming that the spin down is dominated by electromagnetic emission by a magnetic dipole with surface polar cap
field ($B_{p,15}=B_p/10^{15}$ Gauss), the spin-down luminosity would evolve with
time as \citep{1998A&A...333L..87D}
\begin{eqnarray}
L(t)   =  L_0 \frac{1}{(1+t/t_b)^2}
\simeq  \left\{
\begin{array}{ll}
L_0, & t \ll t_b, \\
L_0 (t/\tau)^{-2}, & t \gg t_b,
\end{array}
\right.
\label{Lt}
\end{eqnarray}
where $ L_0 = 1.0 \times 10^{49}~{\rm erg~s^{-1}} (B_{p,15}^2 P_{0,-3}^{-4} R_6^6)$
is the characteristic spin-down luminosity, and $t_b = 2.05 \times 10^3~{\rm s}~ (I_{45} B_{p,15}^{-2} P_{0,-3}^2 R_6^{-6})$
is the characteristic spin-down time scale.

Using all GRBs showing X-ray plateau phases, \citet{2008MNRAS.391L..79D} discovered a tight correlation between $L_0$ and $t_b$ (Dainotti relation).
Subsequently, the Dainotti relation has been used to measure cosmological parameters			 \citep{2009MNRAS.400..775C,2010MNRAS.408.1181C,2013MNRAS.436...82D,2014ApJ...783..126P,2015A&A...582A.115I}.
However, the cosmological constraints are loose. The main reason is that the sample is not properly selected, which induced a large intrinsic scatter on the correlation. In previous works, all GRBs with X-ray plateaus
are used to derive the Dainotti relation, and then constrain cosmological parameters. However, a new-born magnetar can be spun down through
a combination of electromagnetic dipole and gravitational wave quadrupole emission \citep{1986bhwd.book.....S}. The X-ray luminosity of GRBs is given by the energy input from electromagnetic
and gravitational wave into the surrounding medium \cite{1998A&A...333L..87D,2001ApJ...552L..35Z,2011MNRAS.413.2031M}. Therefore, in order to standardize GRBs as standard candles through Dainotti relation, only X-ray plateaus caused by the same physical mechanism (electromagnetic dipole radiation or gravitational wave) can be used. Similar as supernova cosmology, only type Ia supernovae from accretion channels can be treated as standard candles. In this paper, we perform a first attempt to standardize long GRBs with X-ray plateaus dominated by electromagnetic dipole radiations as reliable standard candles.

The structure of this paper is arranged as follows. In next section, the GRB sample is given. In section 3, we show the Dainotti relation. We standardize GRBs using the Dainotti relation in section 4. in section 5, the cosmological constraints from the calibrated the Dainotti relation are shown. Summary is given in section 6.

\section{GRB sample}

If the energy injection from electromagnetic dipole emission of millisecond magnetars is larger than the external shock emission,
the light curves of X-ray afterglow show a plateau with a constant luminosity followed by a decay index about $-2$ \citep{1998A&A...333L..87D,2001ApJ...552L..35Z}. This afterglow
behavior is clean and independent of the complex physics of external shock emission, such as the
fraction of energy going into the electrons, the magnetic field, the shocked electrons and the
surrounding medium \citep{2006RPPh...69.2259M}. Below we adopt this
plateau phase to standardize long GRBs.

The long GRB sample is selected from the total Swift GRBs up to July 2020. The corresponding XRT data is downloaded from the UK Swift Science Data Centre (https://www.swift.ac.uk/xrt$_{-}$curves/). The data processing is given in Refs \citep{2007A&A...469..379E,2009MNRAS.397.1177E,2010A&A...519A.102E}. All the well-sampled X-ray afterglows have a plateau phase with a constant luminosity followed by a decay index of about $-2$ in the X-ray afterglow light
curves. This behavior is well predicted by energy injection from the rotational energy from the newly born magnetars. Until now, there has been a lot of research on the GRB plateau phase \citep{2007ApJ...670..565L,2007ApJ...662.1093W,2013MNRAS.430.1061R,2014ApJ...785...74L,2015ApJ...805...89L,2019ApJ...883...97Z}. The main finding is that magnetars are central engine for GRBs with plateau phases. The selected GRBs in our sample are divided
into two groups (Gold and Silver samples) according to the behaviors of the XRT light curves in the 0.3-10 keV.
The Gold sample is selected in terms of the following five criteria.
\begin{itemize}
	\item There is an obvious plateau and its slope is strictly zero. In addition, we employ the term $(t_{b} -
	t_{1})/(t_{b})$ to describe the duration of the plateau ($T_{p}$), where $t_{1}$ represents the time of the first data point. In
	general, the value of $T_{p}$ is required to be close to 1.0, which indicates the plateau lasts a considerable time. We require
	this value to be at least greater than 0.75.
	\item The decay phase should span a long time, at least 5$t_b$. The plateau phase is followed by a decay with $t^{-2}$ and the duration of
	decay phase ($T_{\rm decay}$) can be described by a simple function $(t_{\rm last} - t_{b})/t_{b}$, where $t_{\rm last}$ is the
	time of the last point.
	\item There are no weak flares, especially during the plateau.
	\item Enough data is required and the distribution is not clustered. This is to ensure continuity of data points.
	\item The reduced chi-square ($\chi^{2}_{r}$) of fit is close to 1.0, preferably within (0.8, 1.5).
\end{itemize}	
The first three criteria are to make sure that the central engine is powered by a newly born magnetar, and play the main role in the
radiation process. The other two criteria are the improvement of the confidence of the fits. There are 10 long GRBs in the Gold
sample, within the redshift range (1.45, 4.65).
The Silver sample consists of GRBs that exhibit an expected plateau followed by a decay with $t^{-2}$. Some bursts of this sample
do not have an obvious plateau, or a few data points in the plateau, but an expected one may exist combining the BAT data. There
are 21 GRBs in this sample. The maximum redshift is 5.91. All bursts fit well with the energy injection model. The $\chi^{2}_{r}$
is between 0.82 and 1.84.

\begin{figure*}	
	\centering
	\includegraphics[width=0.31\hsize]{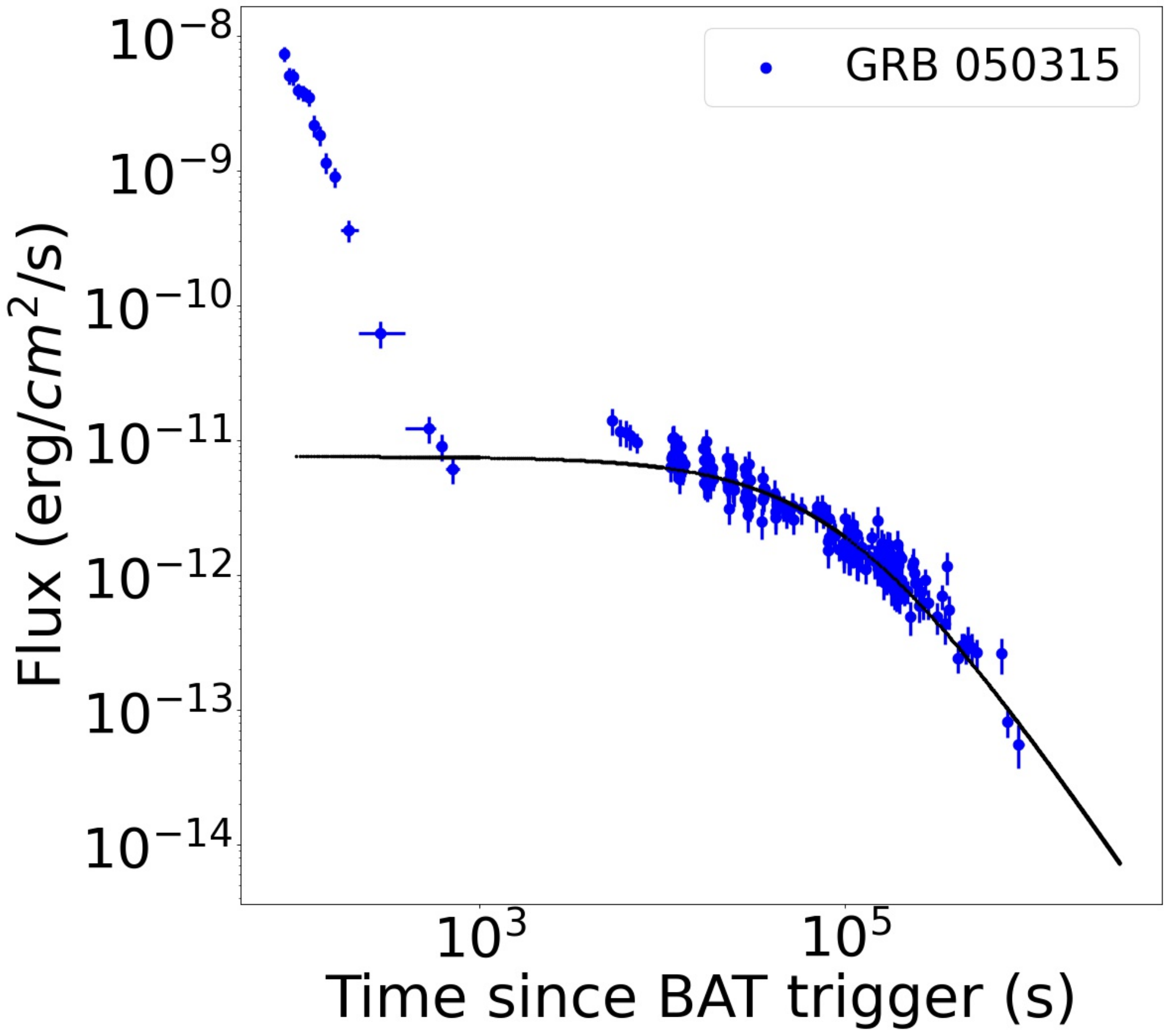}
	\includegraphics[width=0.31\hsize]{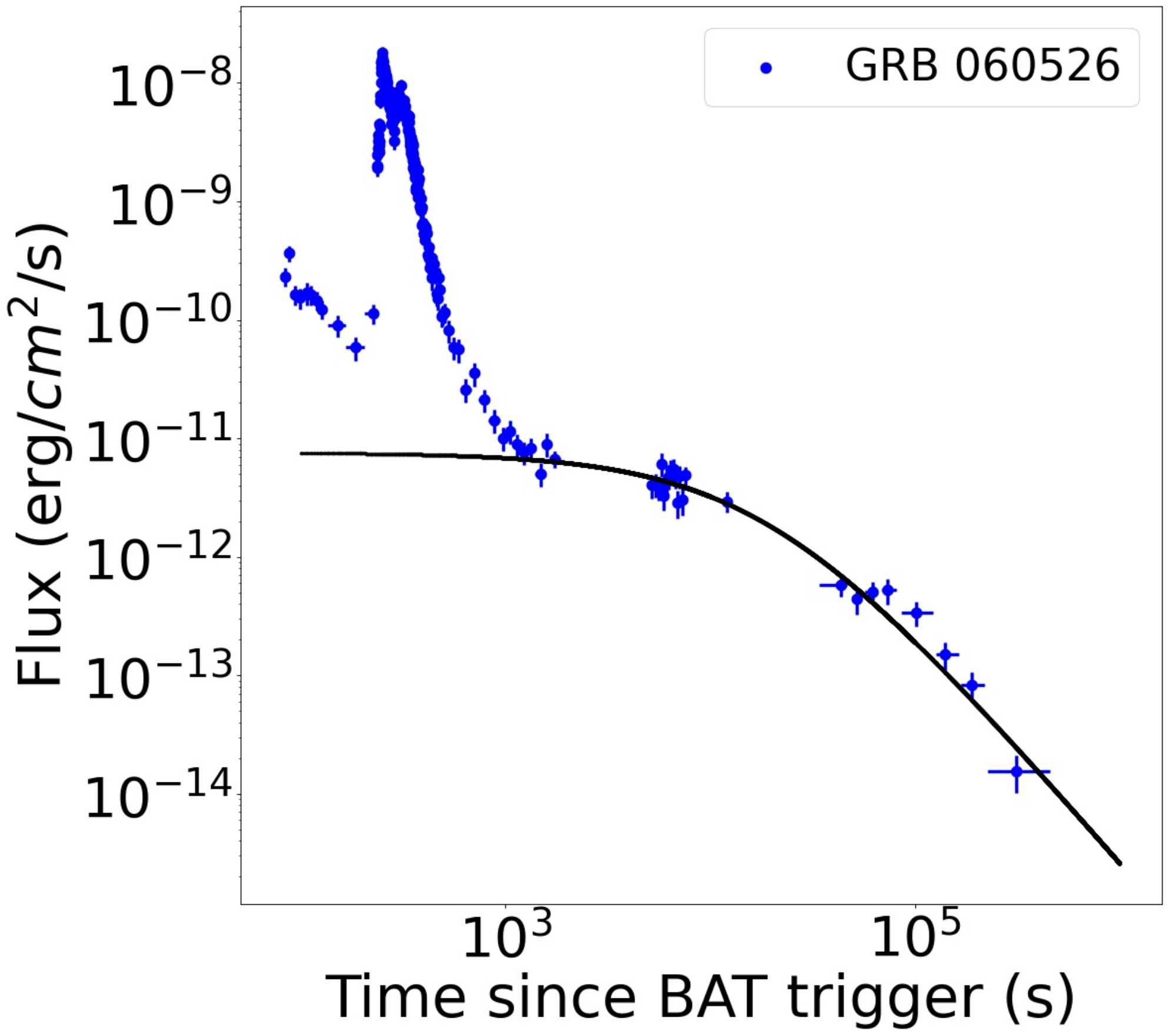}
	\includegraphics[width=0.31\hsize]{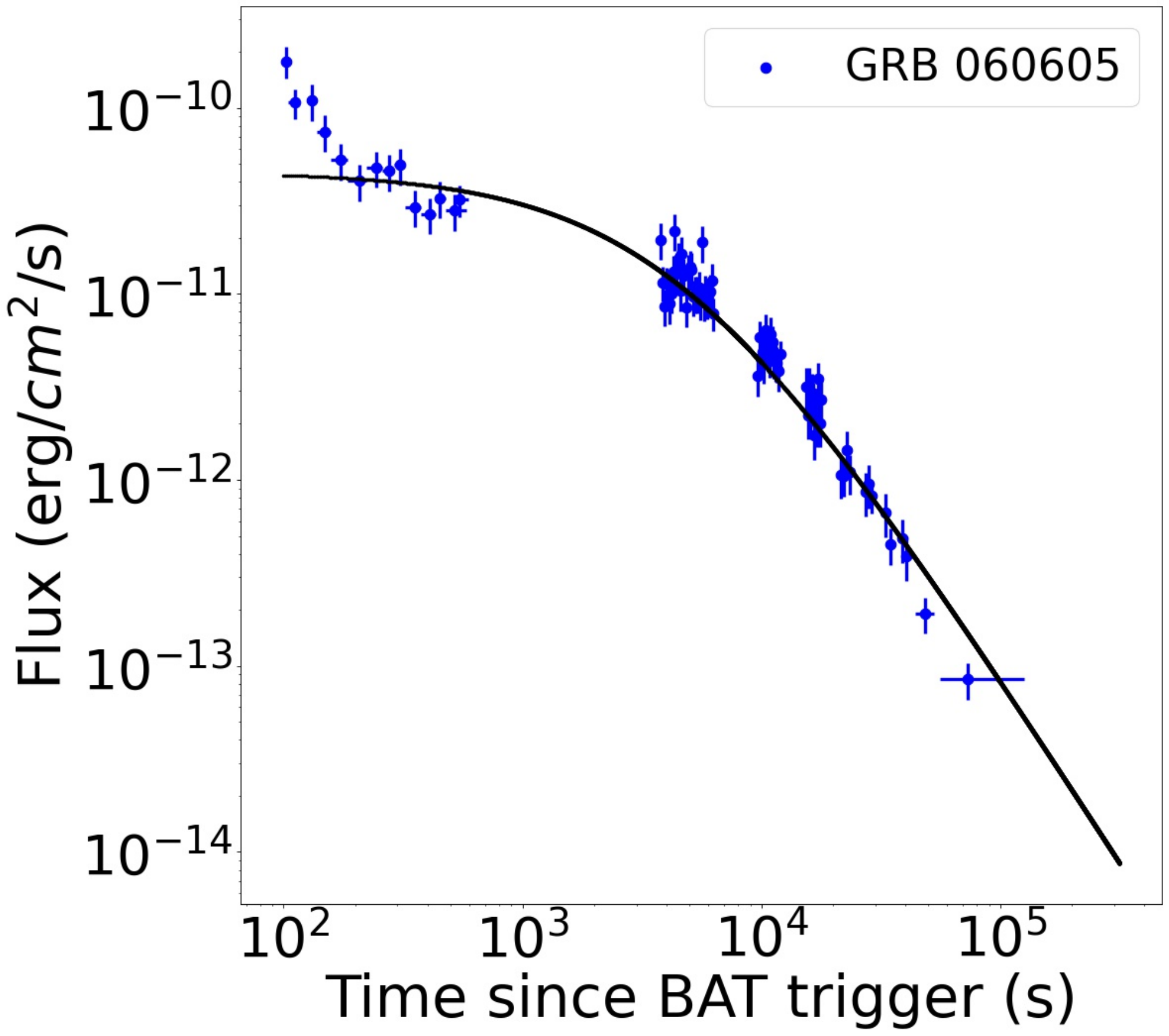}
	\includegraphics[width=0.31\hsize]{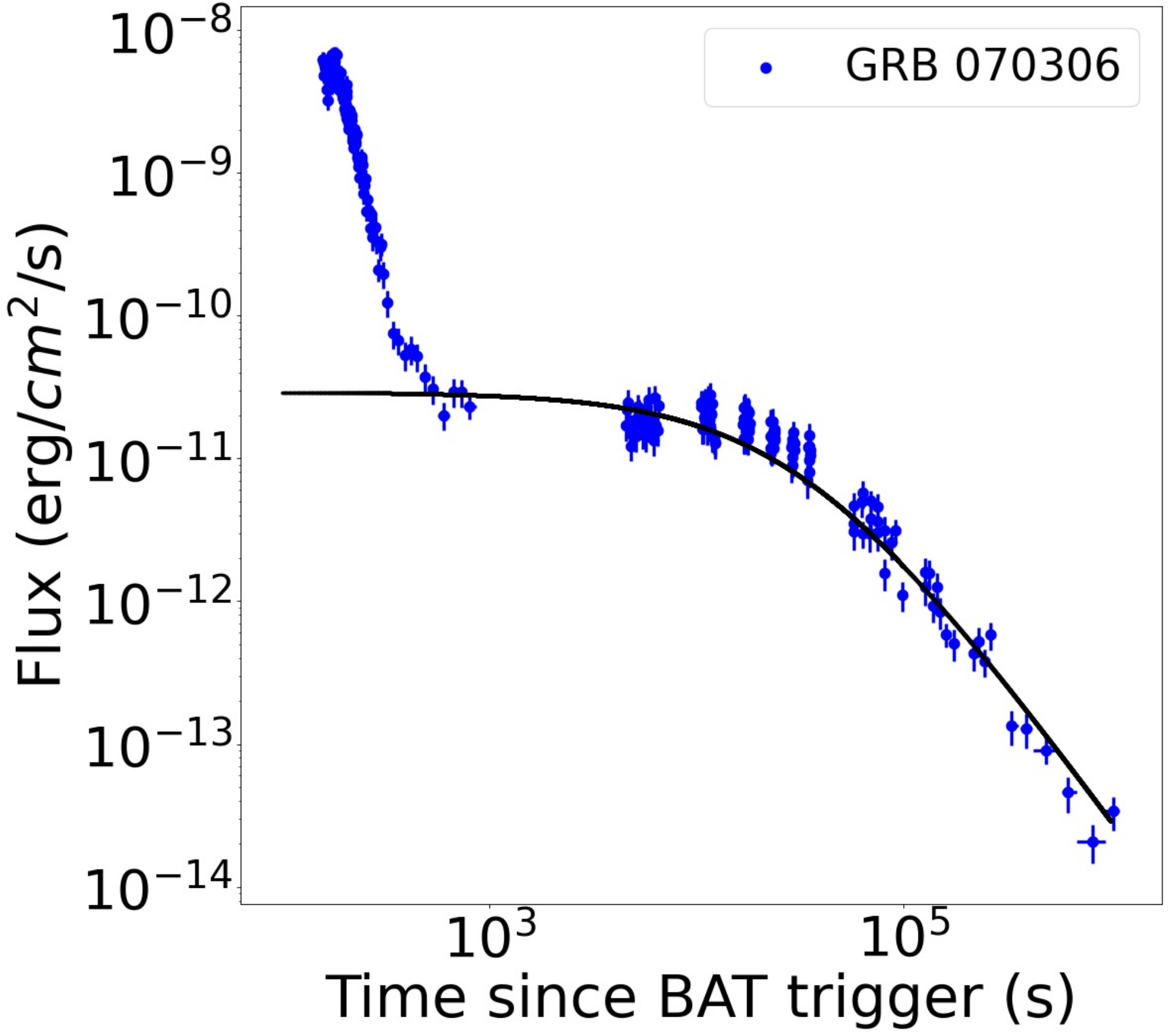}
	\includegraphics[width=0.31\hsize]{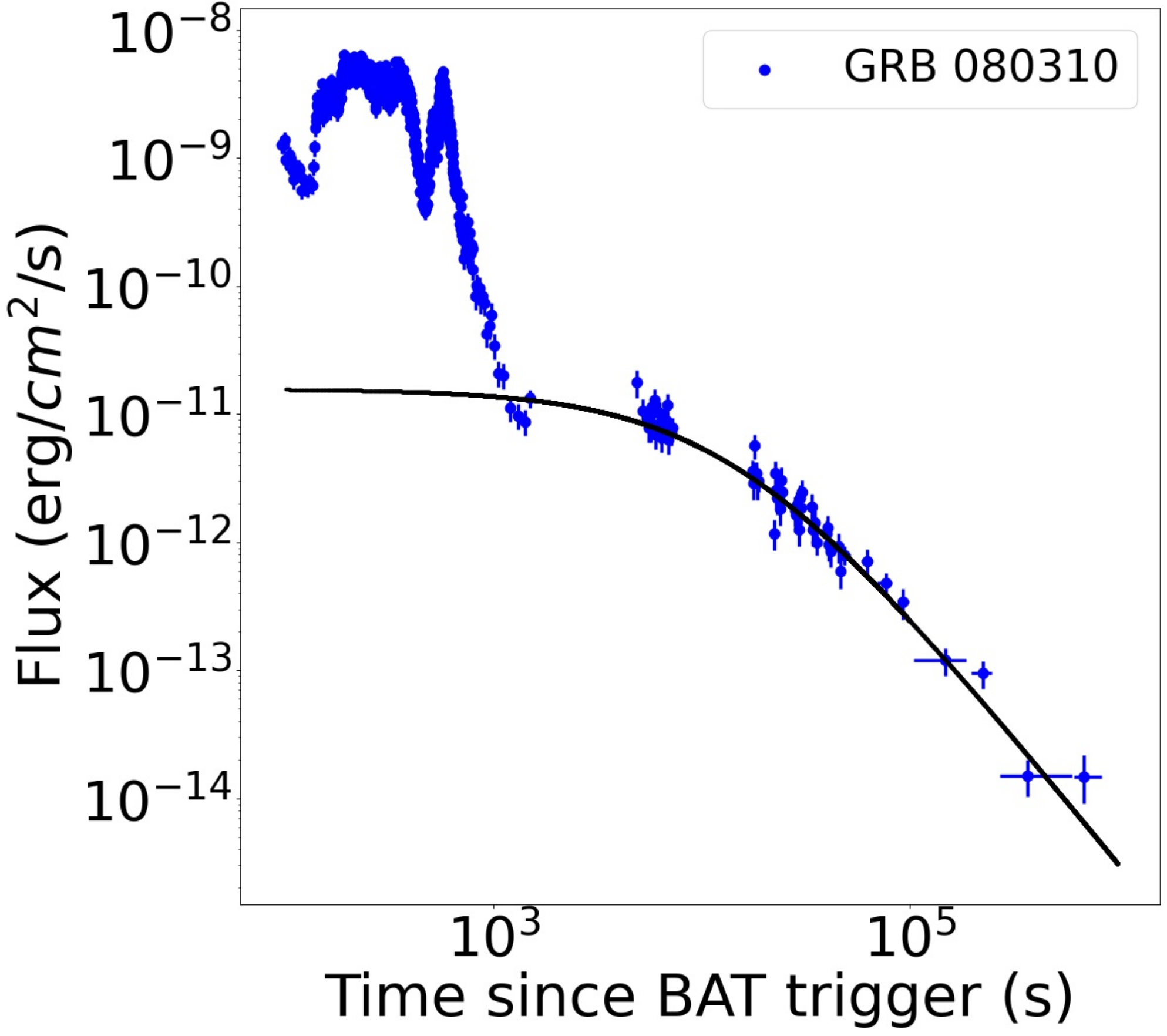}
	\includegraphics[width=0.31\hsize]{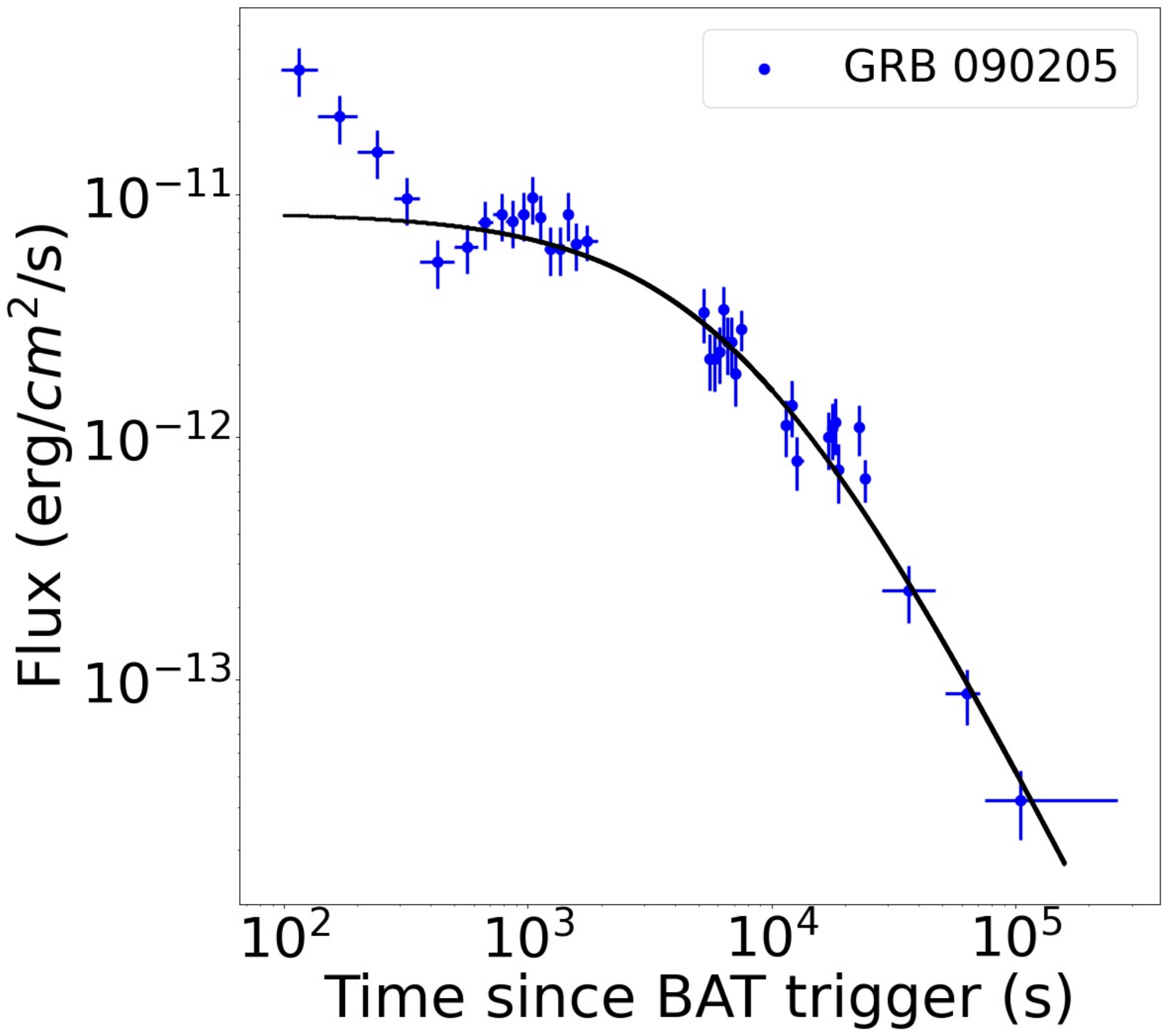}
	\includegraphics[width=0.31\hsize]{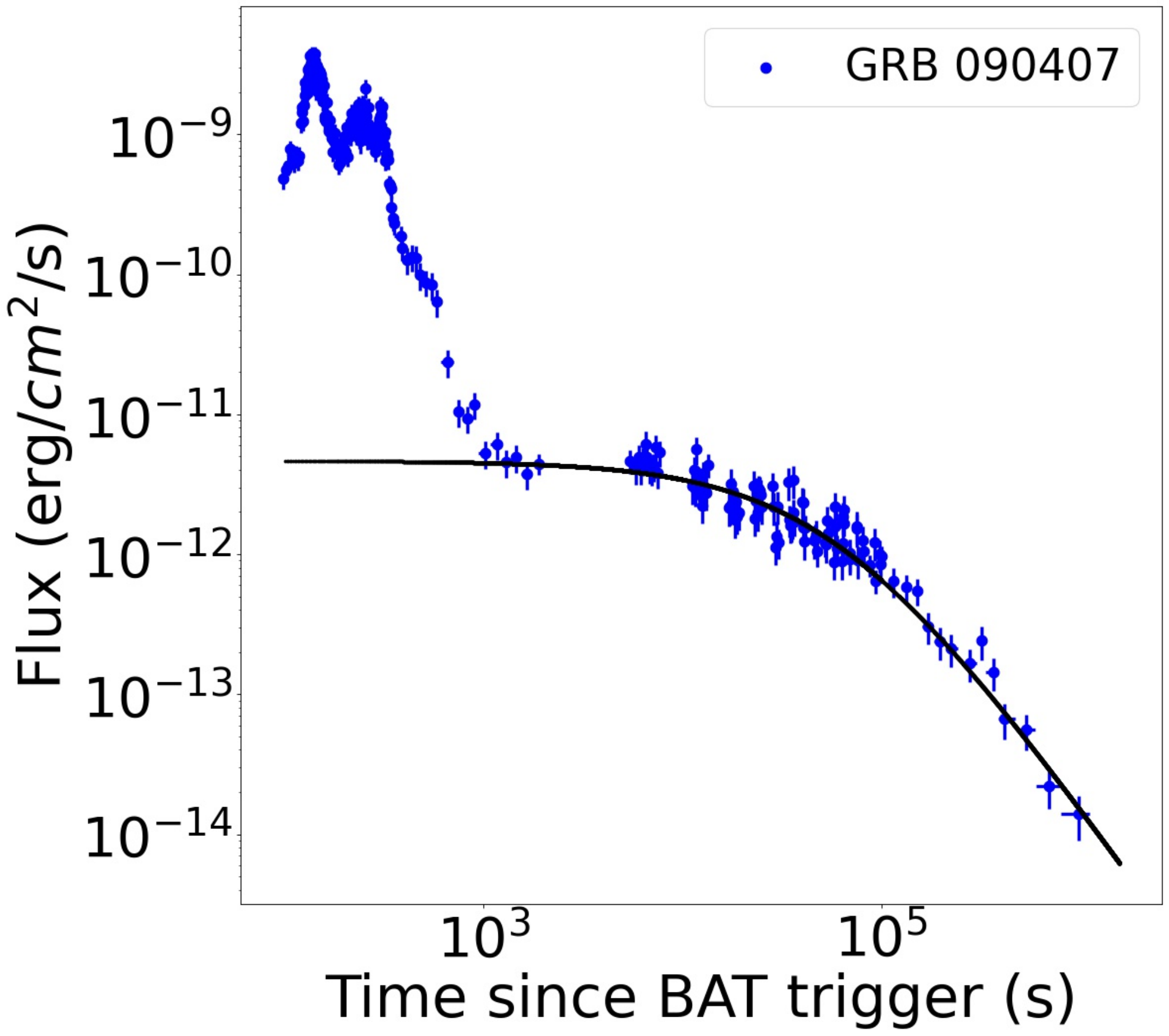}
	\includegraphics[width=0.31\hsize]{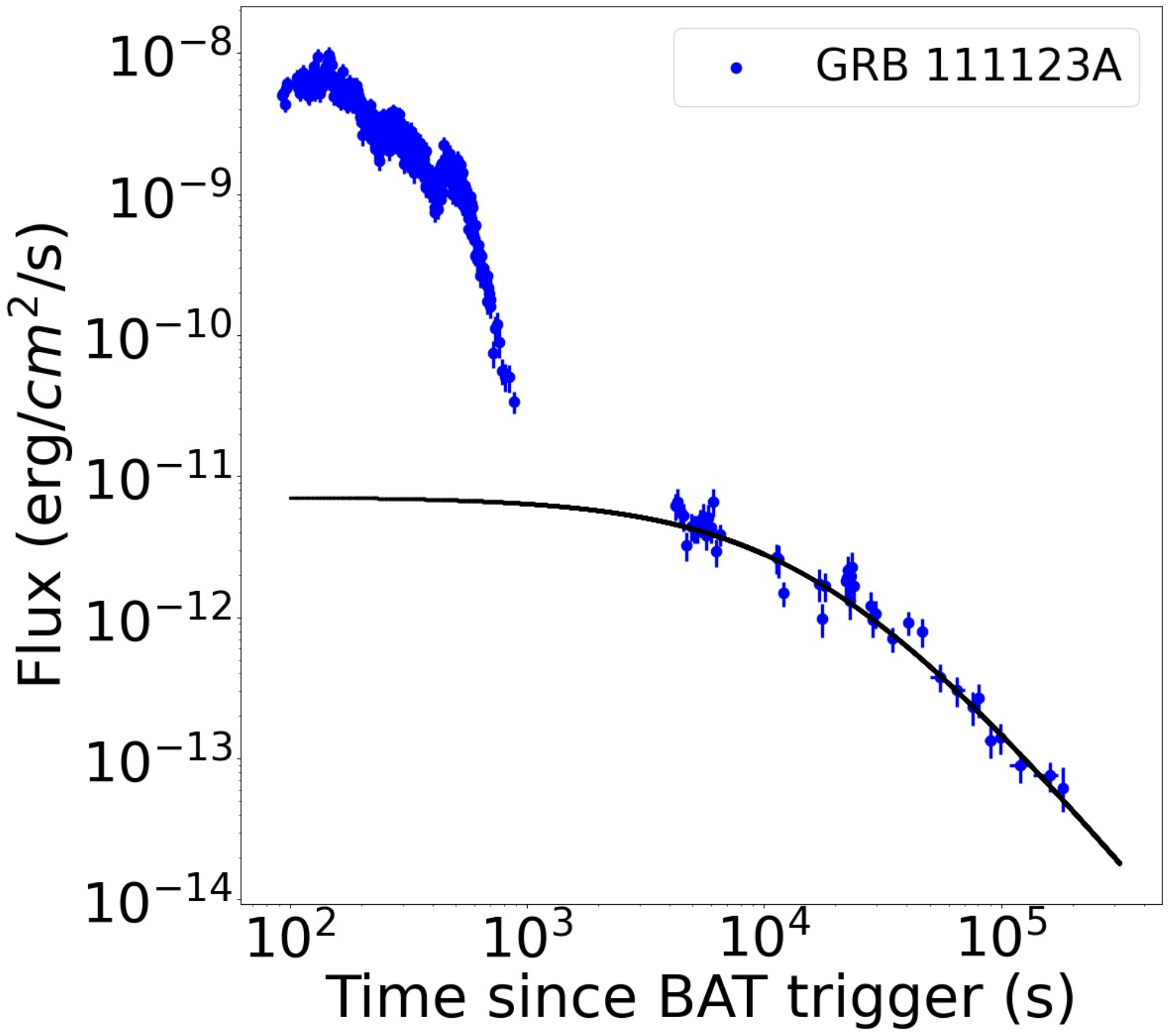}
	\includegraphics[width=0.31\hsize]{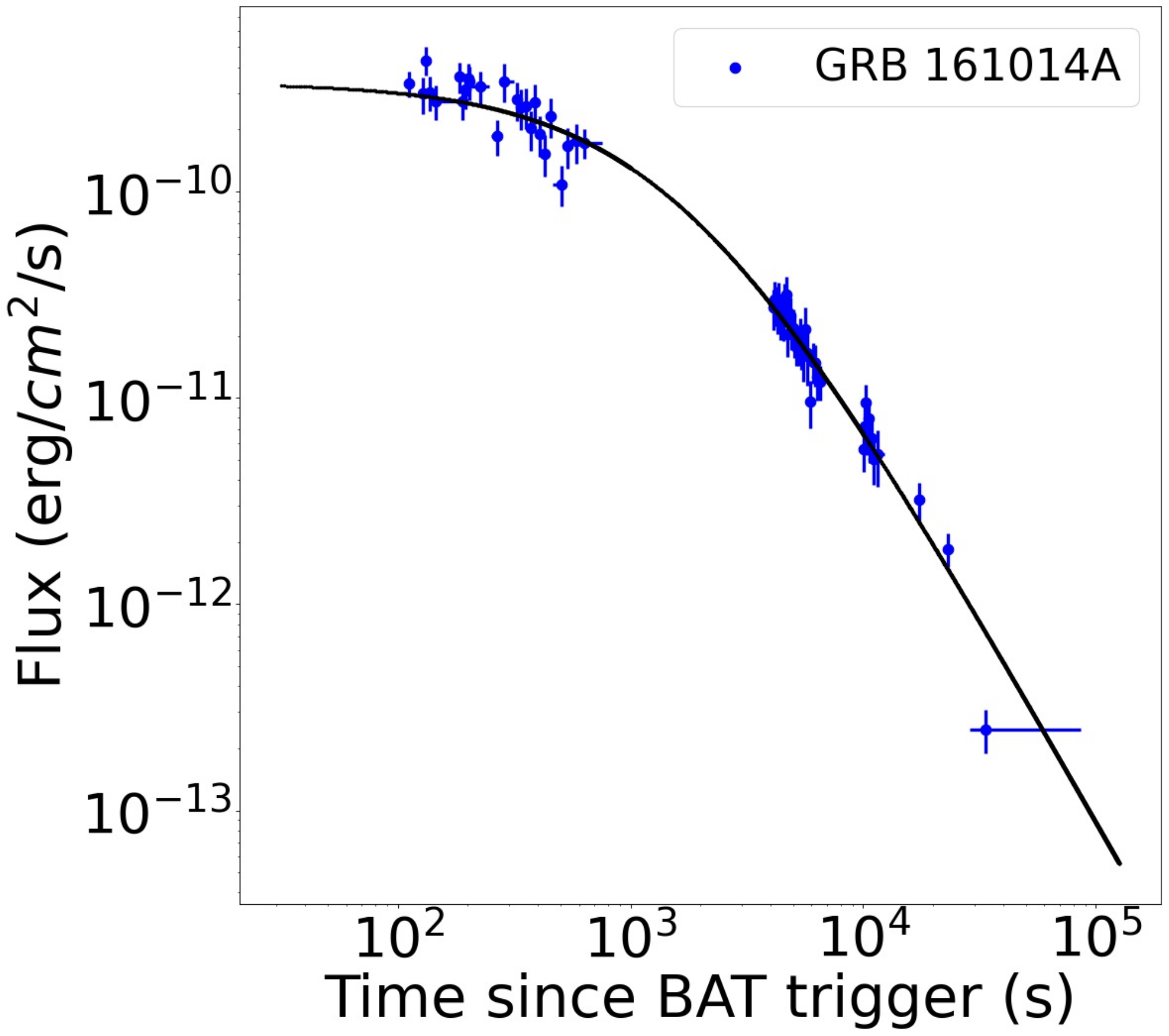}
	\includegraphics[width=0.31\hsize]{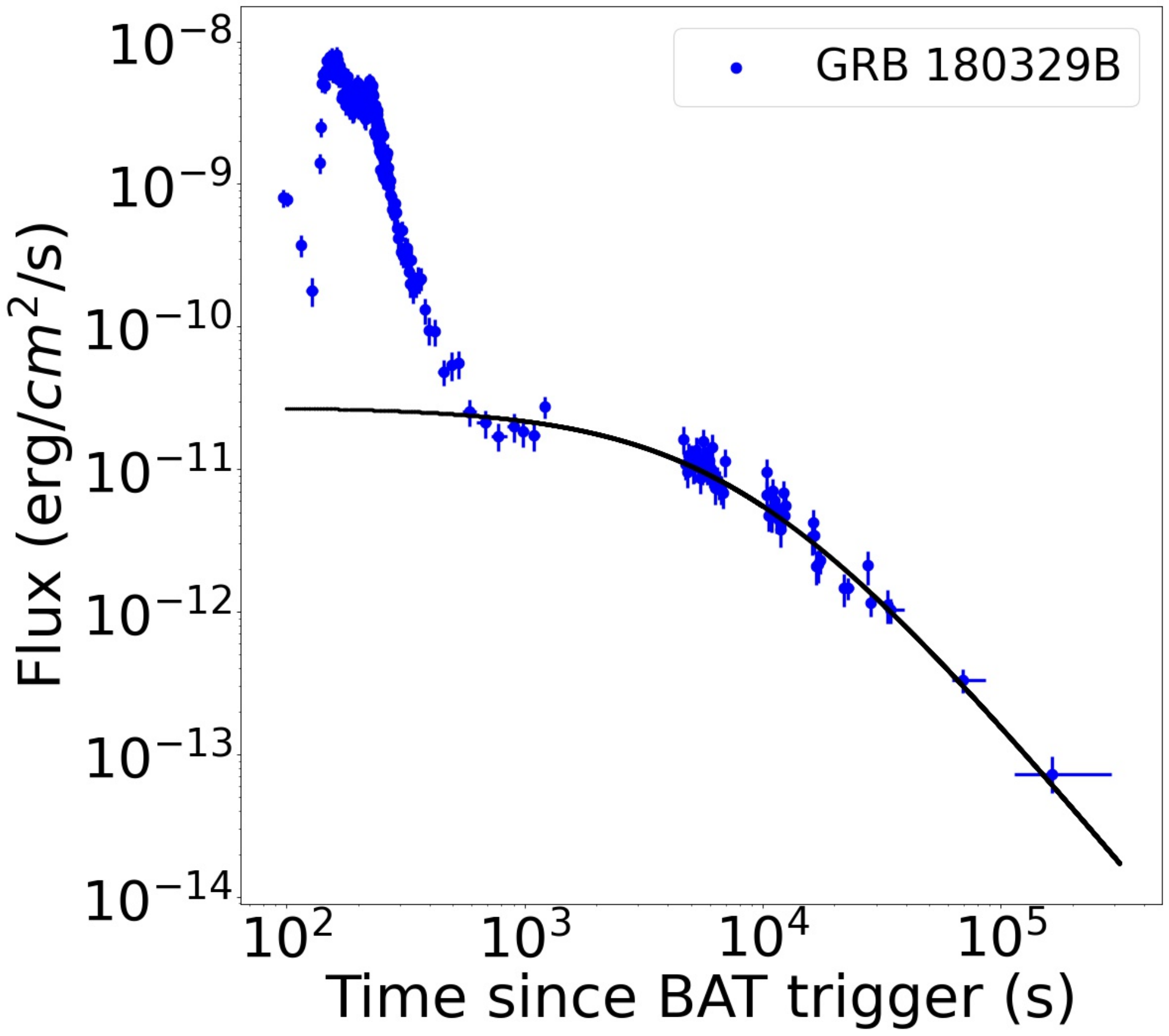}
	\caption{The XRT light curves (0.3-10 keV) of the GRBs in Gold sample. The black solid curves are the best fits with a smooth
		power-law model to the data (blue points). }
	\label{gold}
\end{figure*}

The XRT light curves of the Gold and Silver samples along with the broken power-law fittings using equation (\ref{Lt}) (black curves) are shown in
Figures \ref{gold} and \ref{silvers1}, respectively. The best-fitting results are summarized in Table \ref{T1}.

\section{Dainotti correlation}

With $F_0$ derived above, the luminosity of plateau phase is \citep{2007ApJ...662.1093W,2008MNRAS.391L..79D,2010ApJ...722L.215D,2011ApJ...730..135D}
\begin{equation}
L_0 =  4\pi d_{L}^{2}F_0/(1+z)^{1-\beta},
\end{equation}
where $z$ is the redshift and $\beta$ is the spectral index in the plateau phase. The term $(1+z)^{1-\beta}$ is used to perform
the K-correction \citep{2001AJ....121.2879B}, which converts
the luminosity to the 0.3-10 keV range in the rest frame of GRBs. Because we focus on the X-ray light curves, the luminosity in the rest frame range 0.3-10 keV is considered. The values of $z$ and $\beta=\gamma-1$ are listed in Table \ref{T1} for all GRBs.  A flat $\Lambda$CDM model with $\Omega_{m}$ = 0.3 and
$H_{0}$ = 70 km/s/Mpc is assumed.

The correlation between $L_0$ and $t_{b}$ reads as
\begin{equation}
	\label{eq:logL}
	\log \left (\frac{L_0}{10^{47}~\rm erg/s} \right) = k\times \log \frac{t_{b}}{10^3(1+z)~ \rm s}  + b.
\end{equation}
The Dainotti relation can be expressed as $y = kx + b$.
The corresponding likelihood function is \citep{2005physics..11182D,2008MNRAS.391L..79D,2011MNRAS.415.3423W}
\begin{eqnarray}
	\label{eq:likelihood}
	L(k,b,\sigma _{{\mathop{\rm int}} } ) &\propto&
	\prod\limits_i {\frac{1}{{\sqrt {\sigma ^2 _{{\mathop{\rm int}} }  +
					\sigma ^2 _{y_i }  + k^2 \sigma ^2 _{x_i } } }}} \nonumber \\
	&\times&
	\exp [ - \frac{{(y_i - kx_i - b )^2 }}{{2(\sigma ^2 _{{\mathop{\rm
						int}} }  + \sigma ^2 _{y_i }  + k^2 \sigma ^2 _{x_i } )}}].
\end{eqnarray}
The best fitting values of $k$, $b$ and the intrinsic scatter $\sigma_{int}$ are derived by adopting a Bayesian Monte Carlo Markov Chain (MCMC) method with the emcee\footnote{https://emcee.readthedocs.io/en/stable/} package \citep{2013PASP..125..306F}.
In this paper, all fits are performed using this package.

Figure \ref{Fig3} shows the Dainotti relation for the Gold sample. The best fitting results are $k=-1.07_{-0.12}^{+0.13}$ and $b=1.61_{-0.11}^{+0.10}$ with intrinsic scatter $\sigma_{\rm int}=0.22^{+0.08}_{-0.05}$. All errors are expressed in $1\sigma$ range. We only consider the long GRBs with the electromagnetic dipole emissions above the external shock emissions. It is worth noticing that the plateau luminosity is inversely proportional to the timescale of energy injection, supporting that the energy reservoir is almost a constant. This nearly constant energy of newly born magnetars supports that they can be treated as a standard candle, which is similar to that of type Ia supernovae. The Dainotti relation derived from the total (Gold+Silver) sample is  $k= -1.02_{-0.08}^{+0.09}$, $b=1.72_{-0.07}^{+0.07}$. The best fitting parameters are consistent with those of Gold sample. Some selection effects (i.e., the redshift dependence of $t_b$ and $L_0$, the threshold of the detector) would affect the Dainotti relation. Fortunately, this correlation has been tested against selection bias robustly. For example, Dainotti et al. (2013) studied the redshift dependence of $t_b$ and $L_0$ and found this correlation is robust \citep{2013MNRAS.436...82D}. Moreover, after removing the redshift dependence of $t_b$ and $L_0$, the intrinsic slope $b$ was found to be $-1.07_{-0.14}^{+0.09}$ from 101 GRBs \citep{2015mgm..conf.2106D}, which is dramatically consistent with our result. Some works also confirmed this correlation \citep{2010ApJ...722L.215D,2011ApJ...730..135D,2015MNRAS.451.3898D,2016ApJ...828...36D,2017A&A...600A..98D,2019ApJS..245....1T,2019ApJ...883...97Z,2020ApJ...905L..26D}.

\section{Standarized the light curves of GRBs}

We use the Dainotti relation to standardize the afterglow light curves of long GRBs. First, the end time $t_b$ of all
long GRBs are scaled to the same time using $t/t_b$, where $t$ is the observed time. From Eq. (\ref{eq:logL}), the corresponding luminosity $L_0$ can be
acquired. Then, the scaled luminosity $L/L_0$ can be derived, where $L$ is the plateau luminosity fitted from XRT light curves. This method is similar to
standardize type Ia supernovae (SNe Ia) employing Phillips correlation \citep{1993ApJ...413L.105P}. Figure \ref{Fig4} shows the original (left panel) and standardized light curves (right panel) of gold sample. Although the original
light curves of plateaus are diverse, i.e., the luminosity spans more than two orders of magnitude, the scaled light curves show a universal behavior with a dispersion of 0.5 dex for luminosity. This small dispersion supports that the plateau phase can be regarded as a standard candle. We also repeat the same analysis with $H_0$ = 73.5 km/s/Mpc. We found that the value of $b$ becomes smaller, and the value of $k$ is unchanged. Therefore, the standardization is not affected by the value of $H_0$.

\subsection{Calibrating Dainotti Relation}
Due to lack of low-redshift GRBs, many methods have been proposed to calibrate correlations of GRBs \citep{2008A&A...490...31C,2008MNRAS.391L...1K,2008ApJ...685..354L,2011A&A...536A..96W,2016A&A...585A..68W,2019MNRAS.486L..46A}. In this paper, we utilize Hubble parameter data $H(z)$ \citep{2018ApJ...856....3Y}, whose redshift covers (0.07-2.36) to calibrate the Dainotti relation. The calibrated correlation is model-independent.

First, we need to employ the Gaussian process (GP) method to reconstruct a continuous function $h(x)=H(z)$ that is the best representative of a discrete $h(x_{i}) \pm \sigma_{i}$, where $i$ = 1, 2, 3, ..., $N$ and $\sigma_{i}$ is 1$\sigma$ error. The GP method assumes that the value of $h(x_{i})$ at any position $x_{i}$ is random that follows a Gaussian distribution with the expectation $\varepsilon(x)$ and standard deviation $\sigma(x)$.
The expectation and standard deviation are determined from the observational data through a defined covariance function $k(x,x_{i})$ or kernel function (for example, the Matern kernel), and can be given by
\begin{eqnarray}
	\label{eq:gp_mu}
	\mu(x) &=& \sum_{i,j = 1}^{N} k(x, x_{i})(M^{-1})_{ij}f(x_{j}),
\end{eqnarray}
and
\begin{eqnarray}
	\label{eq:gp_sigma}
	\sigma(x) &=& k(x, x_{i}) - \sum_{i,j = 1}^{N} k(x, x_{i})(M^{-1})_{ij}k(x_{j},x),
\end{eqnarray}
where the matrix $M_{ij} = k(x_{i}, x_{j}) + c_{ij}$ and $c_{ij}$ is the covariance matrix of the observed data. For uncorrelated data, the covariance matrix $c_{ij}$ can be simplified as $diag(\sigma^{2}_{i})$. Equations (\ref{eq:gp_mu}) and (\ref{eq:gp_sigma}) specify the posterior distribution of the extrapolated points. For a given data set ($x_{i}, y_{i}$), considering a suitable kernel function $k$($x$, $\tilde{x}$), it is straightforward to calculate the value of function and its covariance. A more detail explanation of GP method can be found in Section 2 of \citet{2012JCAP...06..036S}. In this paper, we use the Mat\'{e}rn kernel which is a usual kernel function. Its form is written as
\begin{eqnarray}
	\label{eq:matern}
	k(x,\tilde{x}) = \sigma^{2}_{f}(1+\frac{\sqrt{3}|x - \tilde{x}|}{l}) \textnormal{exp} (-\frac{\sqrt{3}|x - \tilde{x}|}{l}),
\end{eqnarray}
where, parameters $\sigma_{f}$ and $l$ control the strength of the correlation of the function value and the coherence length of the correlation in $x$, respectively.

Then, the luminosity distance can be rewritten in terms of the Hubble parameter $H(z)$ as
\begin{eqnarray}
\label{eq:NdL}
d_{L}(z) = c(1+z)\int_{0}^{z} \frac{dz'}{H(z')}.
\end{eqnarray}
Making use of the reconstructed function $h(x)$, the values of $H(z)$ can be estimated at different redshifts. The detailed procedure to determine a continuous function $H(z)$ from GP method can be found in \citet{2018ApJ...856....3Y}. Then according to the equation (\ref{eq:NdL}), we have the corresponding luminosity distance, which can be used to fit the parameters $k$ and $b$ of the Dainotti relation. GP Regression can be implemented by taking advantage of the package GaPP \citep{2012JCAP...06..036S} in the Python environment. $H(z)$ data from \citet{2018ApJ...856....3Y} is adopted in the calibration process. There are 37 $H(z)$ data and its range of redshift covers (0.07, 2.36). We could estimate the distances of GRBs with redshifts less than 2.50 using GP method. There are 14 GRBs in this range. Then we achieve the model-independent luminosity distances, which are applied to fit the parameters $k$ and $b$. The best fitting results are
$k=-1.02\pm0.12$ and $b=1.69\pm0.13$, which are consistent with those given by Gold sample in $1\sigma$ confidence level. According to the calibrated Dainotti relation, the luminosity of each GRB can be derived from the observed $t_b$. Then, the luminosity distances of all GRBs in the total sample can be derived model-independently. There have been a lot of works using the Dainotti relation for cosmological purposes \citep{2009MNRAS.400..775C,2010MNRAS.408.1181C,2013MNRAS.436...82D,2014ApJ...783..126P,2015A&A...582A.115I}. In the next section, the calibrated Dainotti relation will be utilized for cosmological constraints. 

\section{Cosmological Constraints}
For the flat $\Lambda$CDM model, the distance modulus is
\begin{equation}
\label{eq:mu}
\mu = 5\log\frac{d_{L}}{\rm {Mpc}} + 25.
\end{equation}
Replacing $d_{L}$ by function $\sqrt{(L_0/(4\pi F_0/(1+z)^{1-\beta}))}$ and combining the calibrated Dainotti
relation, the observed distance modulus and its uncertainty can be derived from
\begin{eqnarray}
\label{eq:muobs}
\mu_{obs} &=& \frac{5}{2} (\log{L_0} - \log \frac{4\pi F_0}{(1+z)^{1-\beta}} - 24.49) + 25,
\end{eqnarray}
and
\begin{eqnarray}
\label{eq:muerr}
\sigma_{obs} &=& \frac{5}{2} ((\log^{2} (\frac{t_{b}}{1+z}) -3) \sigma_{k}^{2} + k^{2}  (\frac{\sigma_{t_{b}}}{t_{b} \ln{10}})^{2} \nonumber \\
&+& \sigma_{b}^{2}+ (\frac{\sigma_{F_0}}{F_0 \ln{10}})^{2} + \sigma_{int}^{2})^{1/2}.
\end{eqnarray}
Here, $\sigma_{int}=0.22$ is the typical systematic error of the Gold sample.
The likelihood function for the parameter $\Omega_m$ can be determined from $\chi^{2}$ statistics,
\begin{eqnarray}
\label{eq:chi}
\chi^{2}(\Omega_m) = \sum_{i=1}^{31}\frac{(\mu_{obs}(z) - \mu_{th}(\Omega_m,z))^{2}}{\sigma_{obs}^{2}},
\end{eqnarray}
where $\mu_{th}(\Omega_{m},z)$ is the theoretical distance modulus calculated from equation (\ref{eq:mu}). From the total 31 GRBs, the best fitting result is $\Omega_m=$ 0.34$\pm 0.05$.

Figure \ref{Fig5} shows the cosmological constraints, Hubble diagram of long GRBs (purple points) and type Ia supernovae (SNe Ia) of the Pantheon sample (blue points) \citep{2018ApJ...859..101S}.
The uncertainty of GRB distance modulus is around 0.6 magnitude. A fit with the flat $\Lambda$CDM model shown as black solid line provides a best-fit
cosmological matter density parameter of $\Omega_m = 0.34\pm 0.05$ (1$\sigma$) in the high redshift range $1.45<z<5.91$, in agreement with the other main cosmological probes \citep{2016A&A...594A..13P,2018ApJ...859..101S}. For a nonflat $\Lambda$CDM model, the constraints are $\Omega_{m}$ = 0.32$_{-0.10}^{+0.05}$ and $\Omega_{\Lambda}$ = 1.10$_{-0.31}^{+0.12}$ (1$\sigma$) shown as a green solid line in Fig. \ref{Fig5} (left panel).
We find that the evidence for nonzero cosmological constant $\Lambda$ from the GRB sample is $3\sigma$.
$\Omega_m$ obtained from GRBs is in agreement with that derived from SNe Ia. However, the constraint on $\Omega_\Lambda$ derived from GRBs is looser than
that from SNe Ia. The reason is that GRBs locate at high-redshift region, where the cosmic expansion is dominated by matter, not dark energy.
\subsection{Testing the $\Lambda$CDM Tension}
Recent work show that high-redshift Hubble diagrams of supernovae, quasars and GRBs deviate from flat $\Lambda$CDM at 4$\sigma$ confidence level \citep{2019A&A...628L...4L}. But a different conclusion was presented \citep{2020MNRAS.492.4456K}. They made a joint analysis of the quasar, $H(z)$ and baryon acoustic oscillation data, and found that the result is consistent with the current spatially-flat $\Lambda$CDM model. We study this deviation using the standardized Hubble diagram of GRBs.
The best fitting value of $\Omega_{m}$ derived from the calibrated Gold+Silver sample is consistent with the $\Lambda$CDM model.
We test the $\Lambda$CDM model adopting our GRB sample and the Pantheon SNe Ia sample \citep{2018ApJ...859..101S}.
The luminosity distance $d_{L}$ can be expanded by the Taylor expansion \citep{2010JCAP...03..005V} in terms of Hubble series parameters (Hubble
constant $H_0$, deceleration $q_0$, jerk $j_0$, snap $s_0$ and lerk $l_{0}$ parameters). They are derived model-independently in the
FLRW metric. Definitions of the cosmographic parameters are
\begin{eqnarray}
\label{eq:q0j0}
H = \frac{\dot{a}}{a}, q = -\frac{1}{H^{2}}\frac{\ddot{a}}{a}, j=\frac{1}{H^3}\frac{\dot{\ddot{a}}}{a},\\ \nonumber
s=\frac{1}{H^4}\frac{\ddot{\ddot{a}}}{a}, l=\frac{1}{H^5}\frac{\dot{\ddot{\ddot{a}}}}{a}\label{j}.
\end{eqnarray}
The luminosity distance can be expanded as a function of $y = z/(1+z)$ in a flat cosmology \citep{2007CQGra..24.5985C,2009A&A...507...53W,2010JCAP...03..005V}
\begin{eqnarray}
\label{eq:Tdl}
d_{L}(y) &=& \frac{c}{H_{0}}(y - \frac{1}{2}(q_{0} - 3)y^{2} + \frac{1}{6}(11-5q_{0} + 3q_{0}^{2} - j_{0})y^{3}\nonumber \\
&+& \frac{1}{24}(50 - 7j_{0}- 26q_{0} + 10q_{0}j_{0} + 21 q_{0}^{2} - 15q_{0}^{3} + s_{0})y^{4} \nonumber \\
&+& \frac{1}{120}(274 - 154q_{0}+141q_{0}^{2}- 135q_{0}^{3} + 105q_{0}^{4} - 47j_{0} \nonumber \\
&+& 10j_{0}^{2}+90q_{0}j_{0}-105q_{0}^{2}j_{0}-15q_{0}s_{0}+9s_{0}-l_{0})y^5 \nonumber \\
&+&\textit{O}(y^6))
\end{eqnarray}
where $H_{0}$, $q_{0}$, $j_{0}$, $s_{0}$ and $l_{0}$ are the current values. The only assumption of the above expansion is the FLRW metric. We
can get the distance modulus from equation (\ref{eq:mu}). The best fitting parameters can be constrained by minimizing
\begin{eqnarray}
\label{eq:chi_qj}
\chi^{2}(H_{0}, q_{0}, j_{0}) &=& \sum_{i=1}^{31}\frac{(\mu_{GRB}(z) - \mu_{th}(z))^{2}}{\sigma_{GRB}^{2}} \nonumber \\
&+&\sum_{j=1}^{1048}\frac{(\mu_{SN}(z) - \mu_{th}(z))^{2}}{\sigma_{SN}^{2}}.
\end{eqnarray}
Even using the series expansion in $y$, the problem of the series truncation remains. The higher the order of the cosmographic
expansion, the more accurate the approximation. But, the more cosmographic parameters, the larger the volume of the parameter
space, and the weaker the constraining strength by degeneracy effects among different parameters \citep{2017A&A...598A.113D}.
We choose the fifth-order expansion to constrain $H_{0}$, $q_{0}$ and $j_0$ by marginalizing $s_0$ and $l_0$ in a large range ($0<s_0,l_0<20$).

In the flat $\Lambda$CDM model,
$q_0=3\Omega_m/2-1$ and $j_0=1$ are expected. We obtained a tight constraint on these parameters from the combined Hubble diagram of
SNe Ia and GRBs with the first three terms of the Taylor expansion. The fitting results, $H_{0}=$ 69.98$_{-0.25}^{+0.48}$ km s$^{-1}$ Mpc$^{-1}$
, $q_{0}=-0.52_{-0.11}^{+0.09}$ and $j_{0}=$ 1.34$_{-0.70}^{+0.81}$, are shown in Figure \ref{Fig6}. In the flat $\Lambda$CDM model, $\Omega_m=0.315\pm0.007$ is given by the final full-mission Planck measurements of CMB \citep{2020A&A...641A...6P}.
Using this value, $q_0=-0.49$ and $j_0=1$ are derived, which is shown as red point in the right panel of Figure \ref{Fig6}. We can see that both the $q_0$
and $j_0$ are consistent with the predictions of  flat $\Lambda$CDM model at high redshifts in $2\sigma$ confidence level. So the tension between
$\Lambda$CDM model and high-redshift GRBs is not as significant as that mentioned in previous work.

\section{Summary}
In this paper, long GRBs with X-ray plateaus dominated by electromagnetic dipole emission ($t^{-2}$) are standardized as a cosmic distance indicator using the Dainotti relation. Compared to the previous research on GRB plateaus, we pay more attention to the different decaying indices after the end of the plateau. The main reason is the different decaying indices represent different physical processes. The scaled light curves of Gold sample have a luminosity dispersion of 0.5 dex. This small dispersion supports that these GRBs can be used as cosmological indicators. The GP method is used to calibrate the Dainotti relation. Using this calibrated correlation, we constrain cosmological parameters, and found that GRB data supports the accelerating universe at $3\sigma$ confidence level. The calibrated GRB Hubble diagram is consistent with $\Lambda$CDM.

In summary, although the number of long GRBs with universal afterglow behavior is small at present. Forthcoming observations by the French-Chinese satellite space-based multi-band astronomical variable objects monitor (SVOM) \citep{Wei2016}, the Einstein Probe (EP) \citep{Yuan2015} and the Transient High-Energy Sky and Early Universe Surveyor (THESEUS) \citep{Amati2018} space missions together with ground- and space-based multi-messenger facilities will allow us to study the poorly explored high-redshift universe.

%% The "ht!" tells LaTeX to put the figure "here" first, at the "top" next
%% and to override the normal way of calculating a float position
%\acknowledgments
\begin{acknowledgments}
We thank the anonymous referee for constructive comments. We thank Bing Zhang and Peter Meszaros for helpful discussions. This work was supported by the National Natural Science
Foundation of China (grant No. U1831207 and 11833003), the Fundamental Research Funds for the Central Universities (No. 0201-14380045), the National Key Research
and Development Program of China (grant No. 2017YFA0402600) and the National SKA Program of China (No. 2020SKA0120300). This work made use of data supplied by the UK Swift Science Data Centre at the University of
Leicester.
\end{acknowledgments}

\bibliography{apj_mg}{}
\bibliographystyle{aasjournal}

\clearpage
%%%%%%%%%%%%%%%%%%%%%%%%%%%%%%%%%%%%%%%%%%%%%%%%%%

%Fig.1

\begin{figure*}
	\centering
	\includegraphics[width=0.31\hsize]{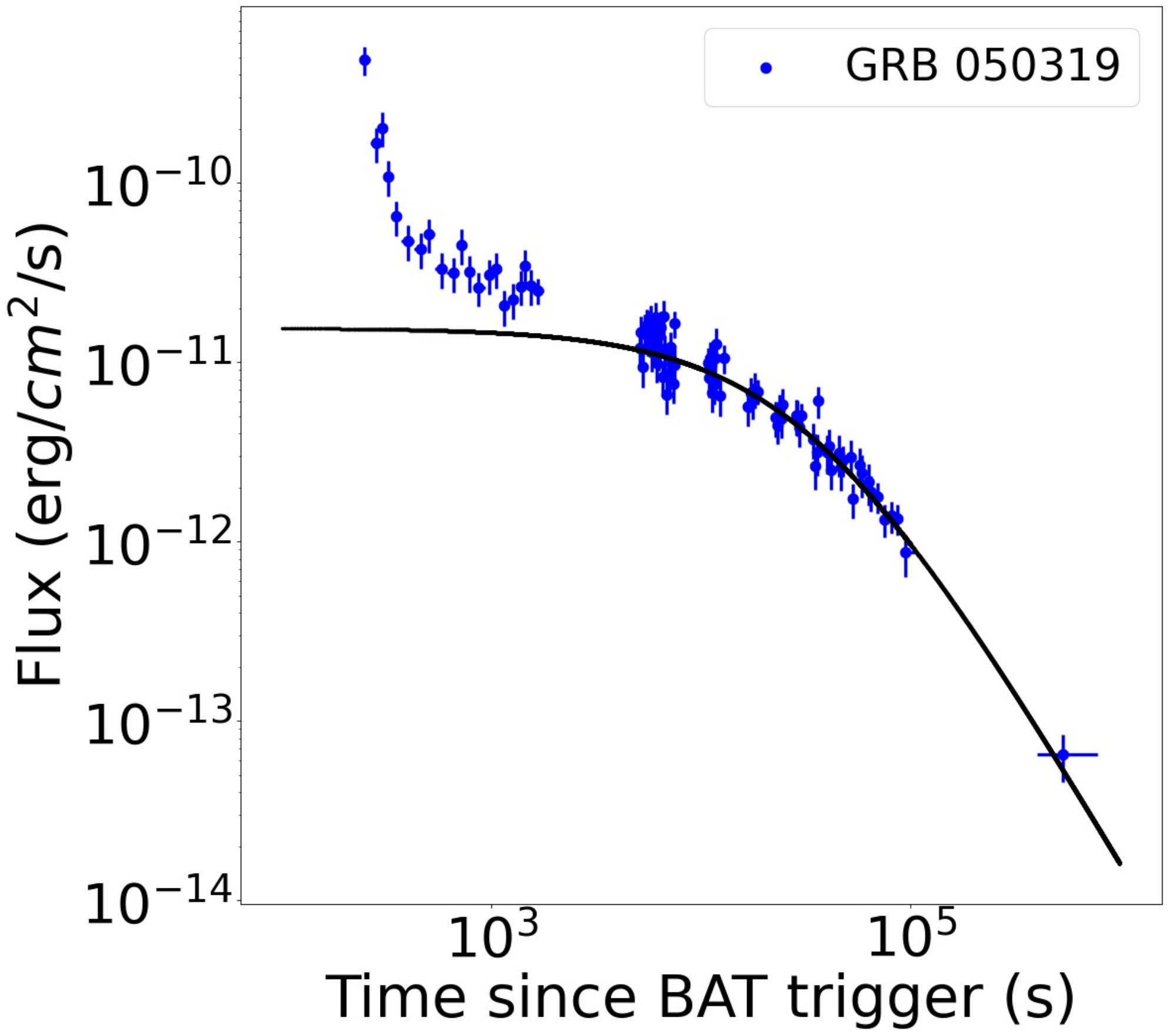}
	\includegraphics[width=0.31\hsize]{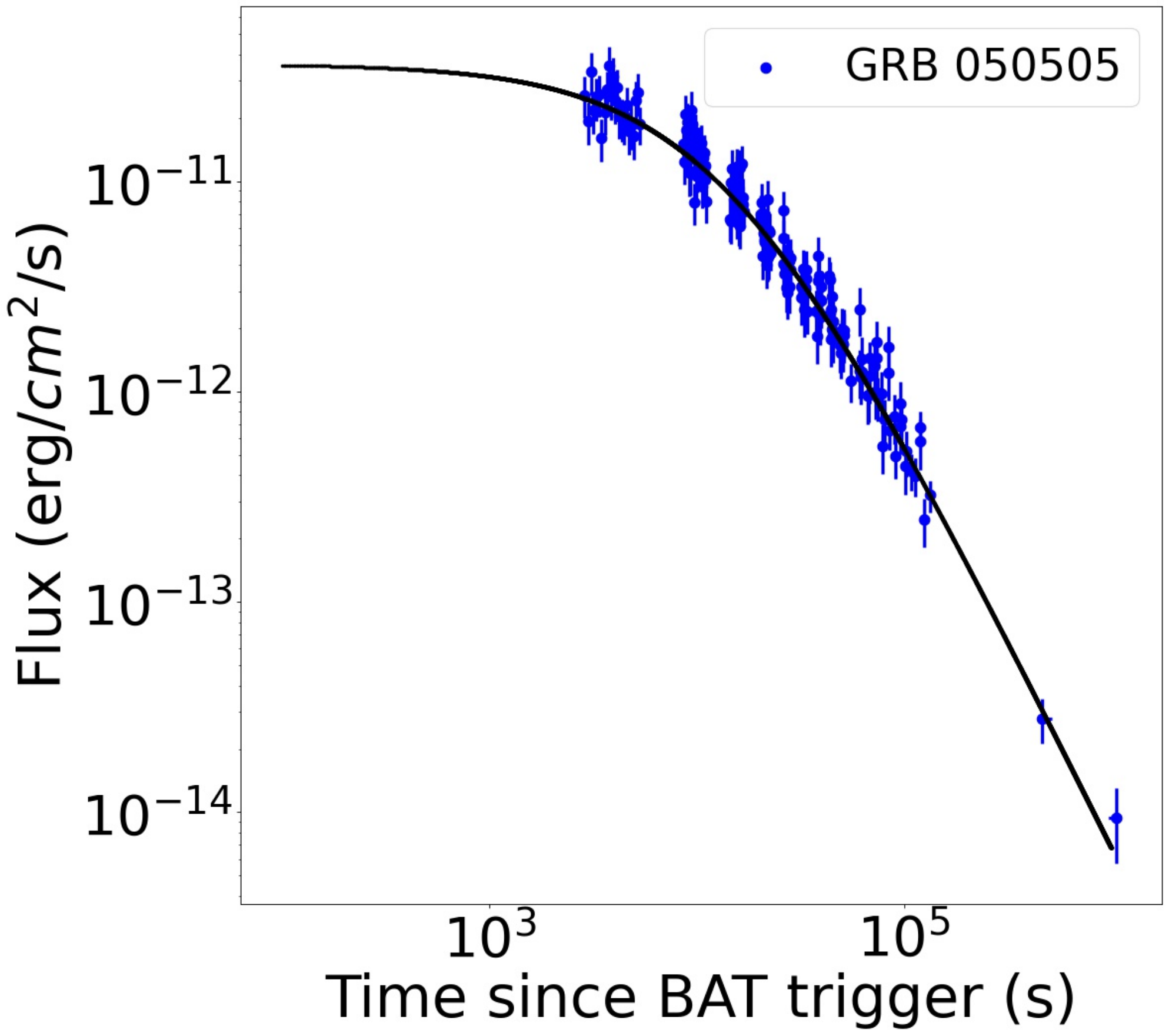}
	\includegraphics[width=0.31\hsize]{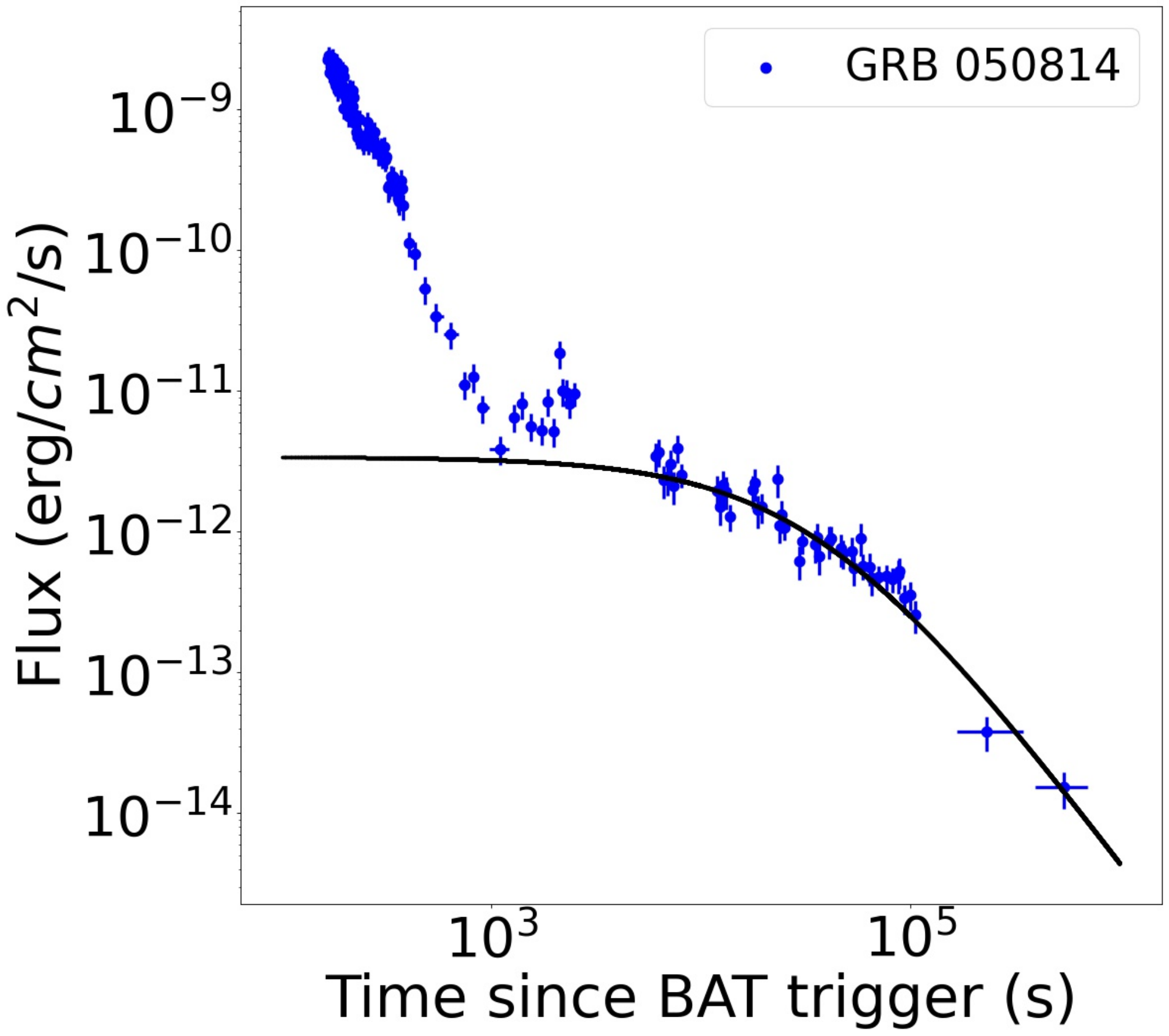}
	\includegraphics[width=0.31\hsize]{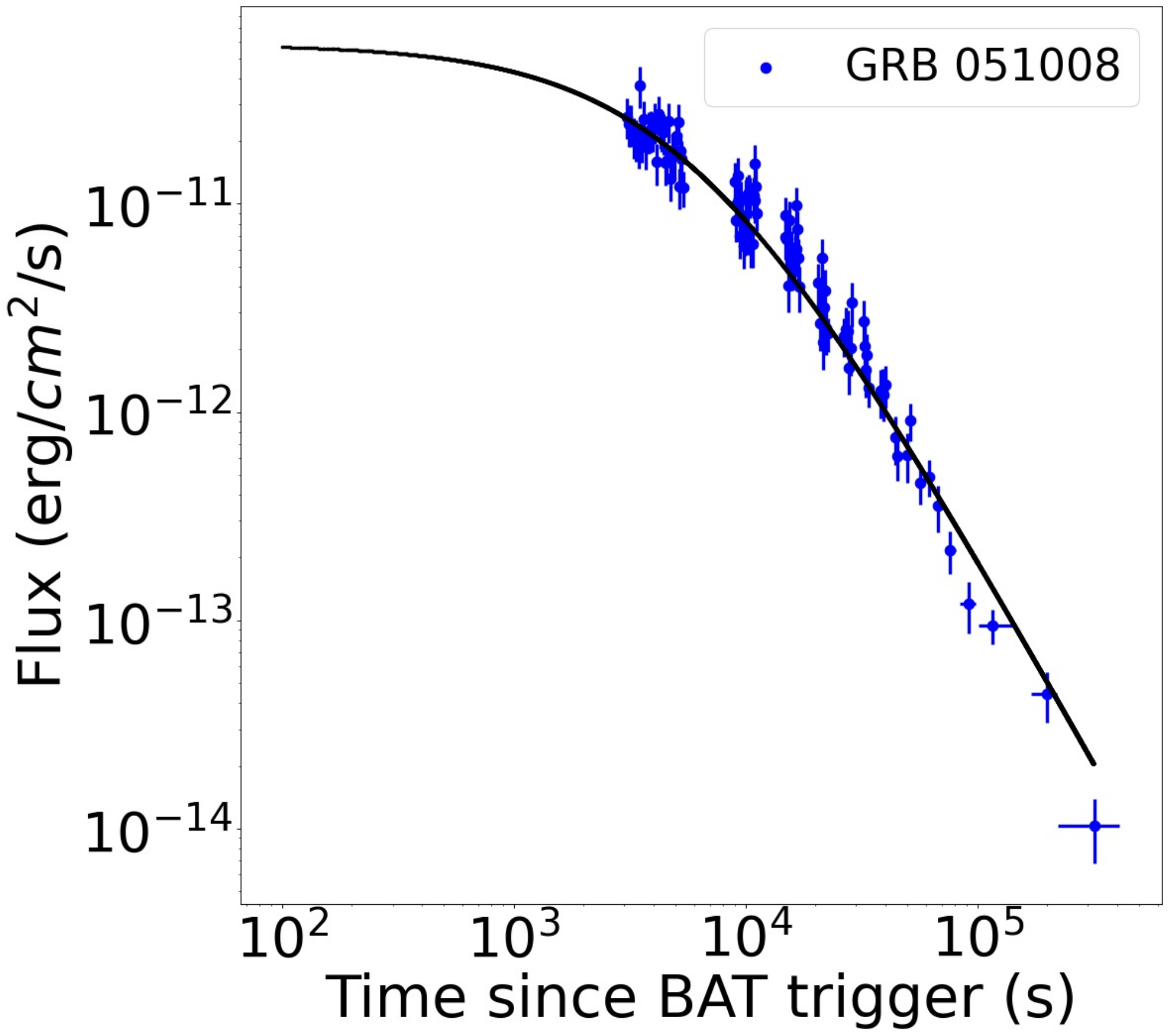}
	\includegraphics[width=0.31\hsize]{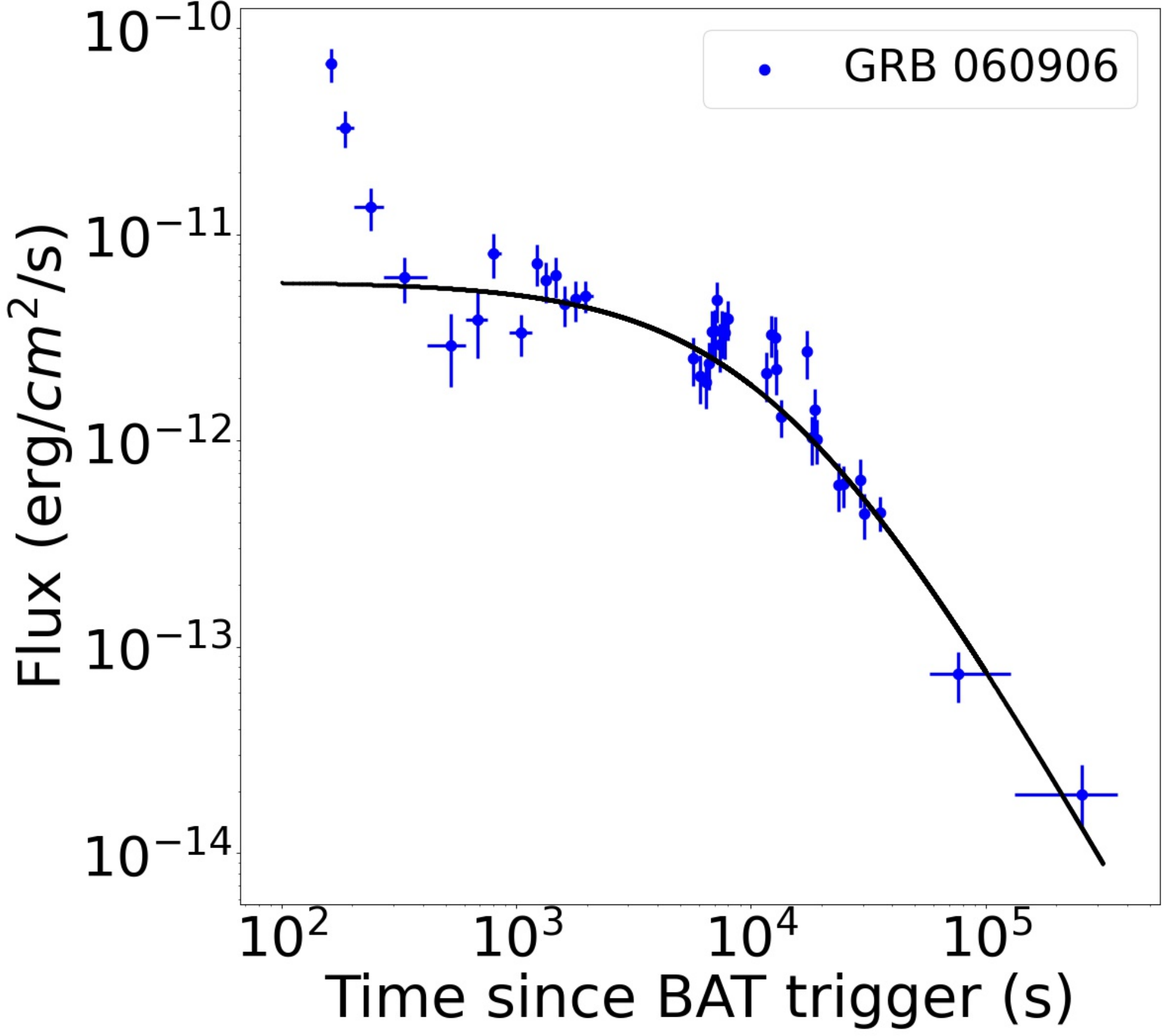}
	\includegraphics[width=0.31\hsize]{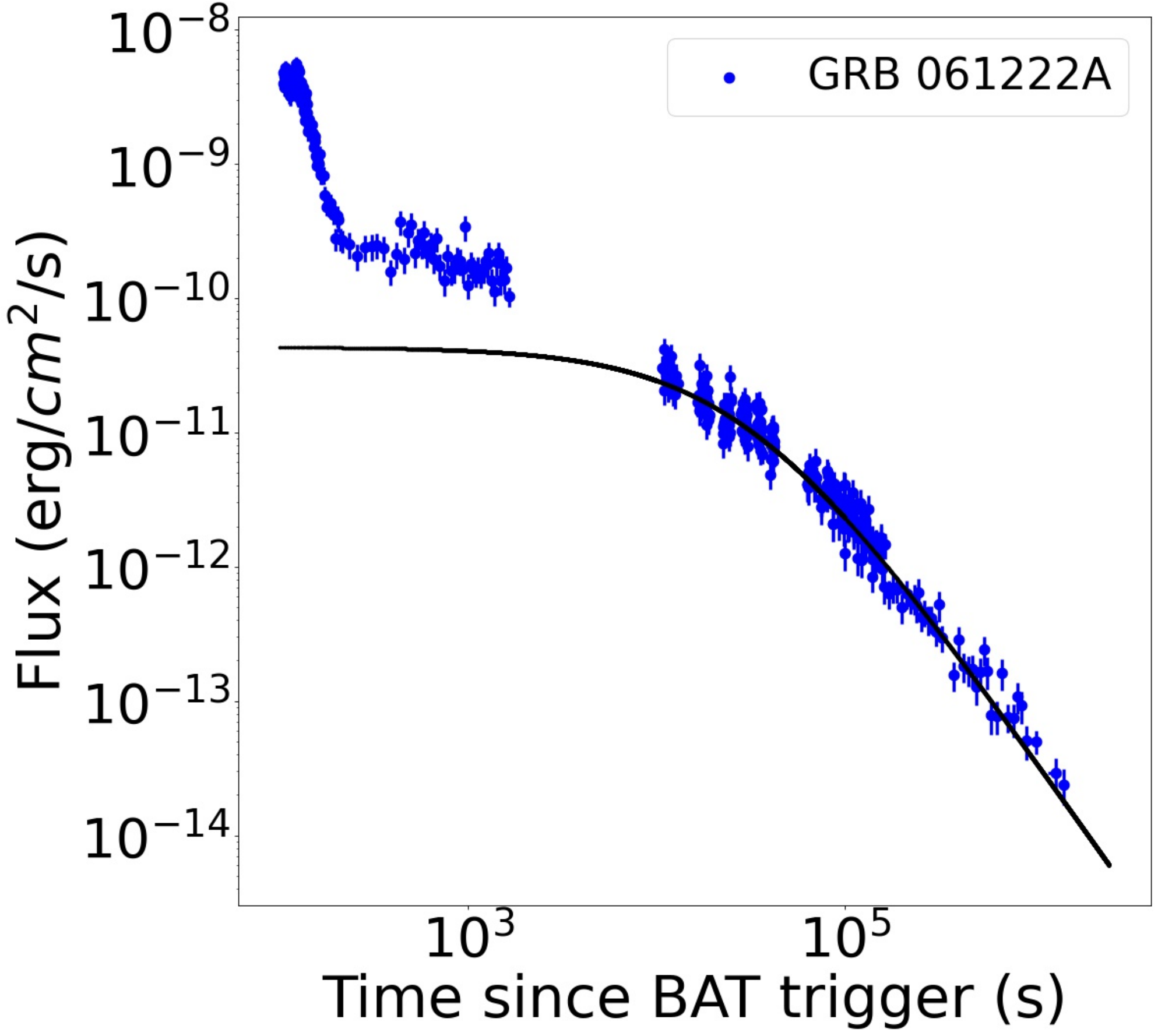}
	\includegraphics[width=0.31\hsize]{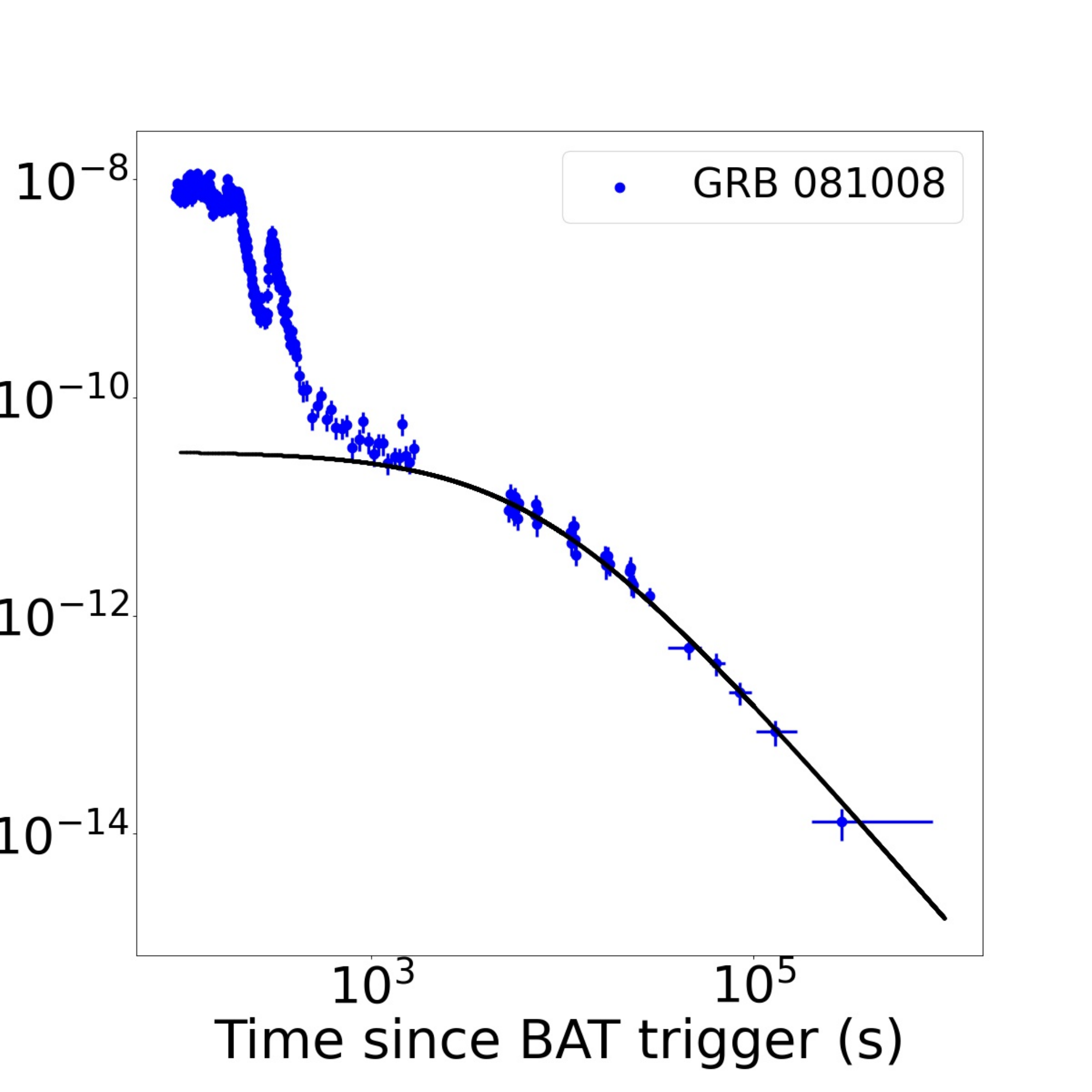}
	\includegraphics[width=0.31\hsize]{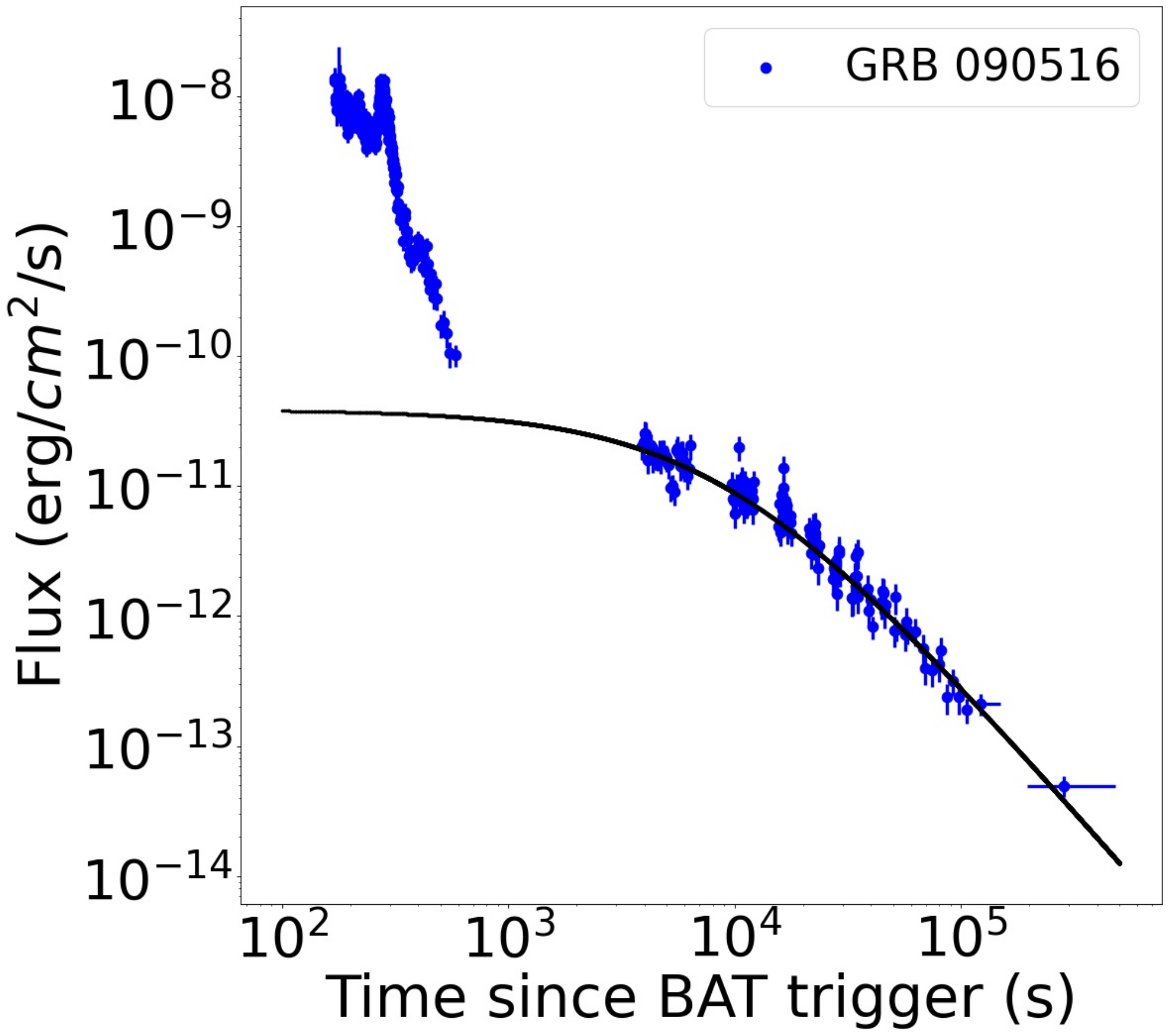}
	\includegraphics[width=0.31\hsize]{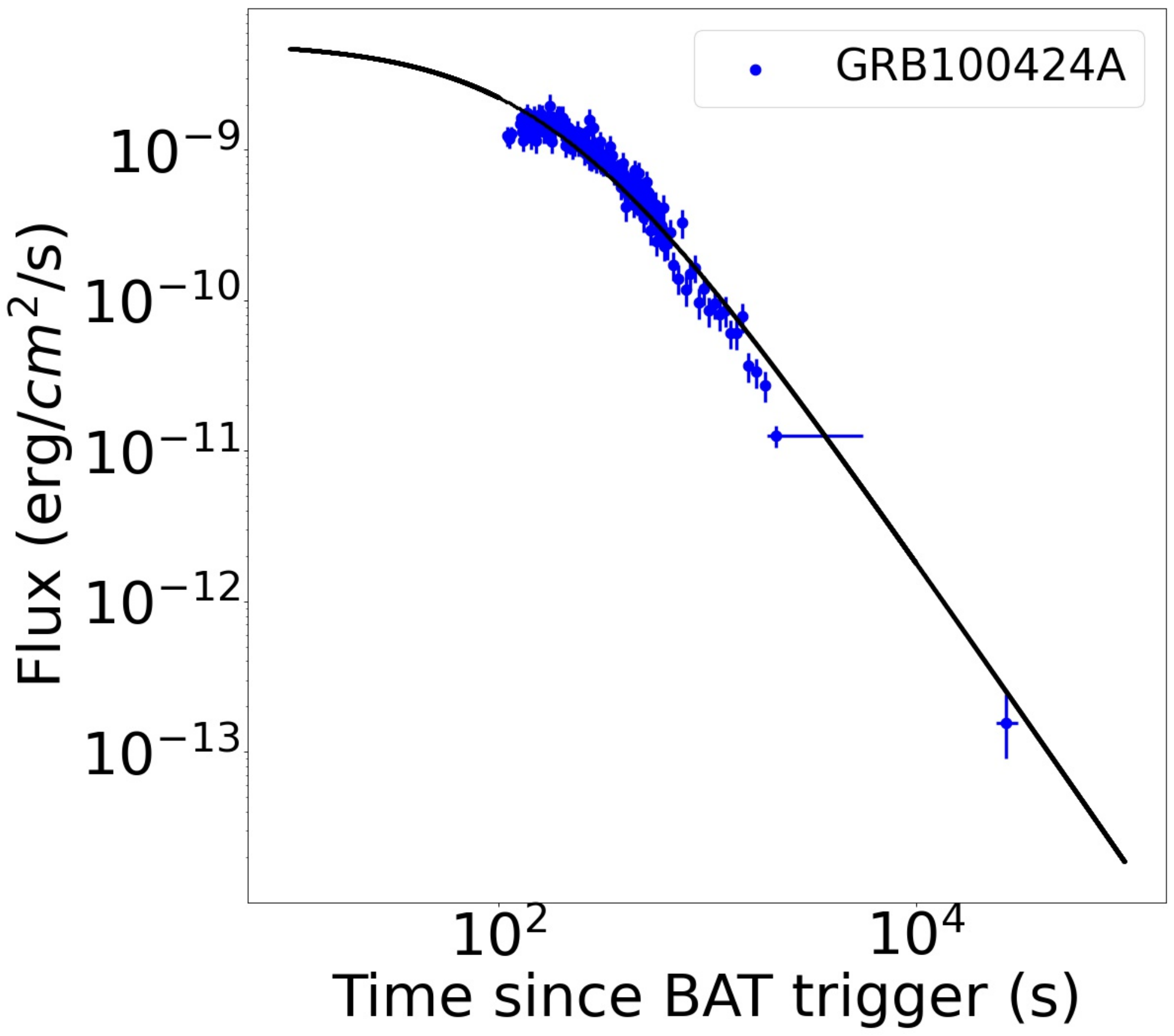}
	\includegraphics[width=0.31\hsize]{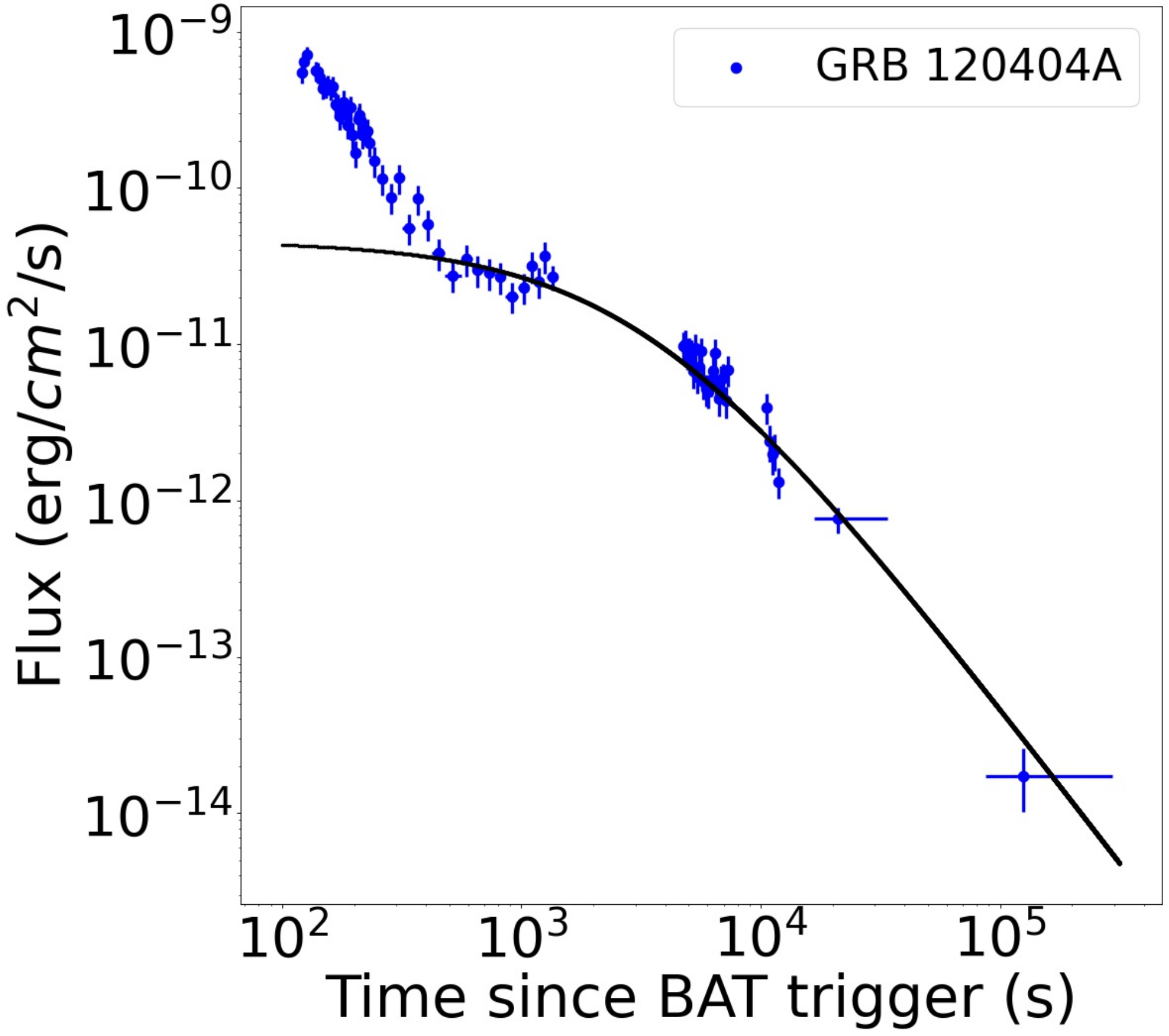}
	\includegraphics[width=0.31\hsize]{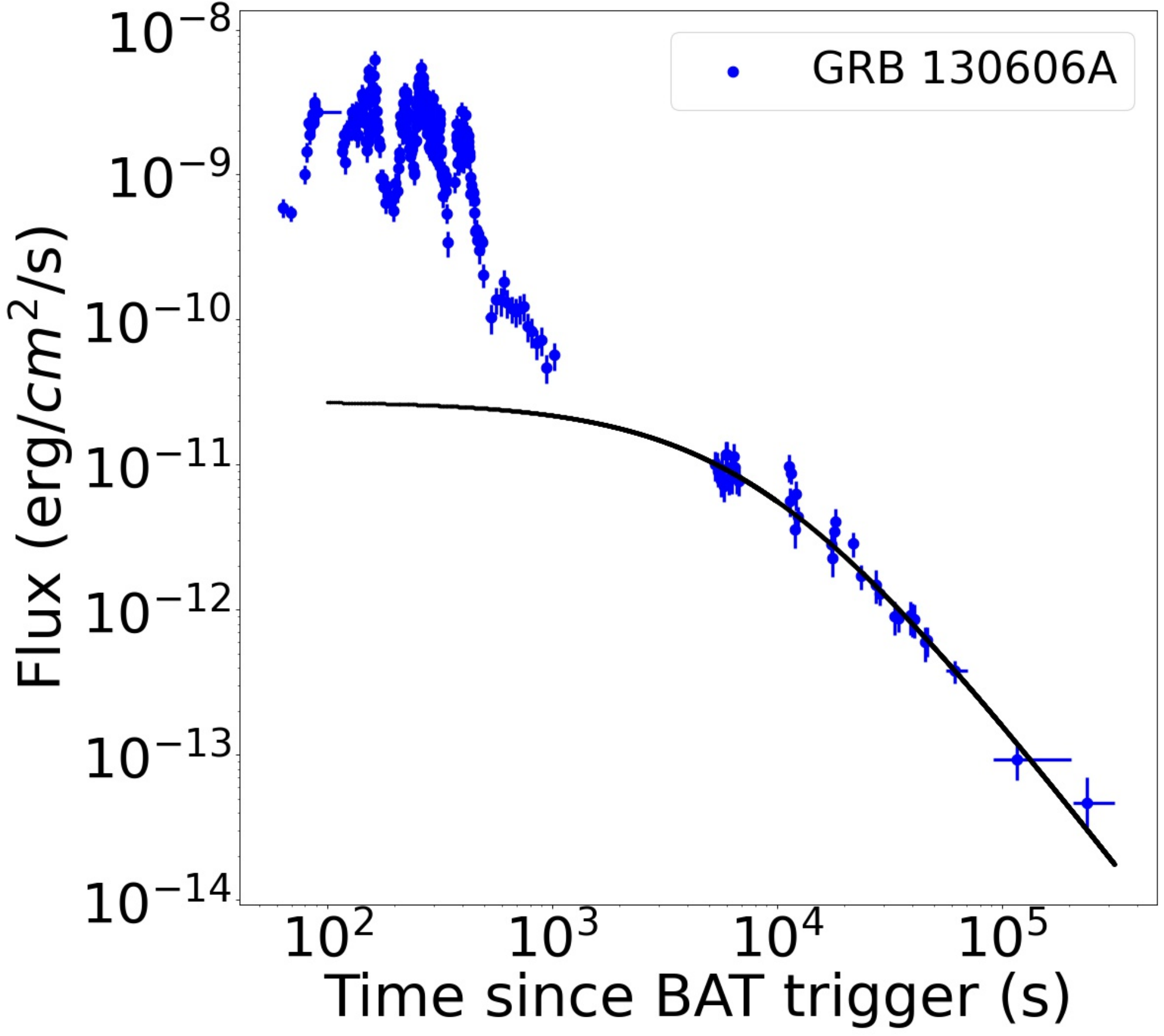}
	\includegraphics[width=0.31\hsize]{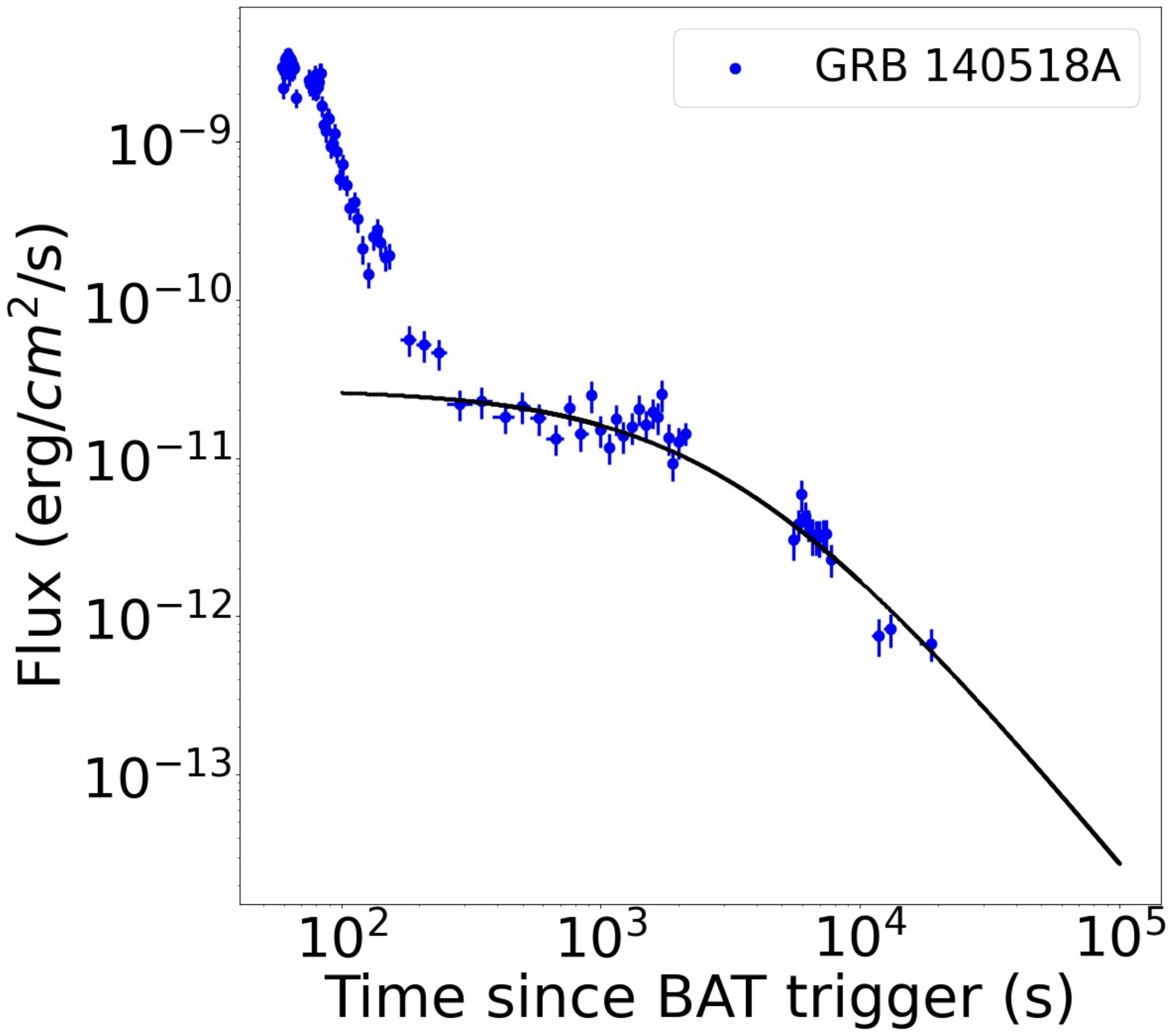}
	\caption{The XRT light curves (0.3-10 keV) of the GRBs in Silver sample. The black solid curves are the best fits with a smooth power-law model to the data (blue points). }
	\label{silvers1}
\end{figure*}

\addtocounter{figure}{-1}
\begin{figure*}
	\centering
	\includegraphics[width=0.31\hsize]{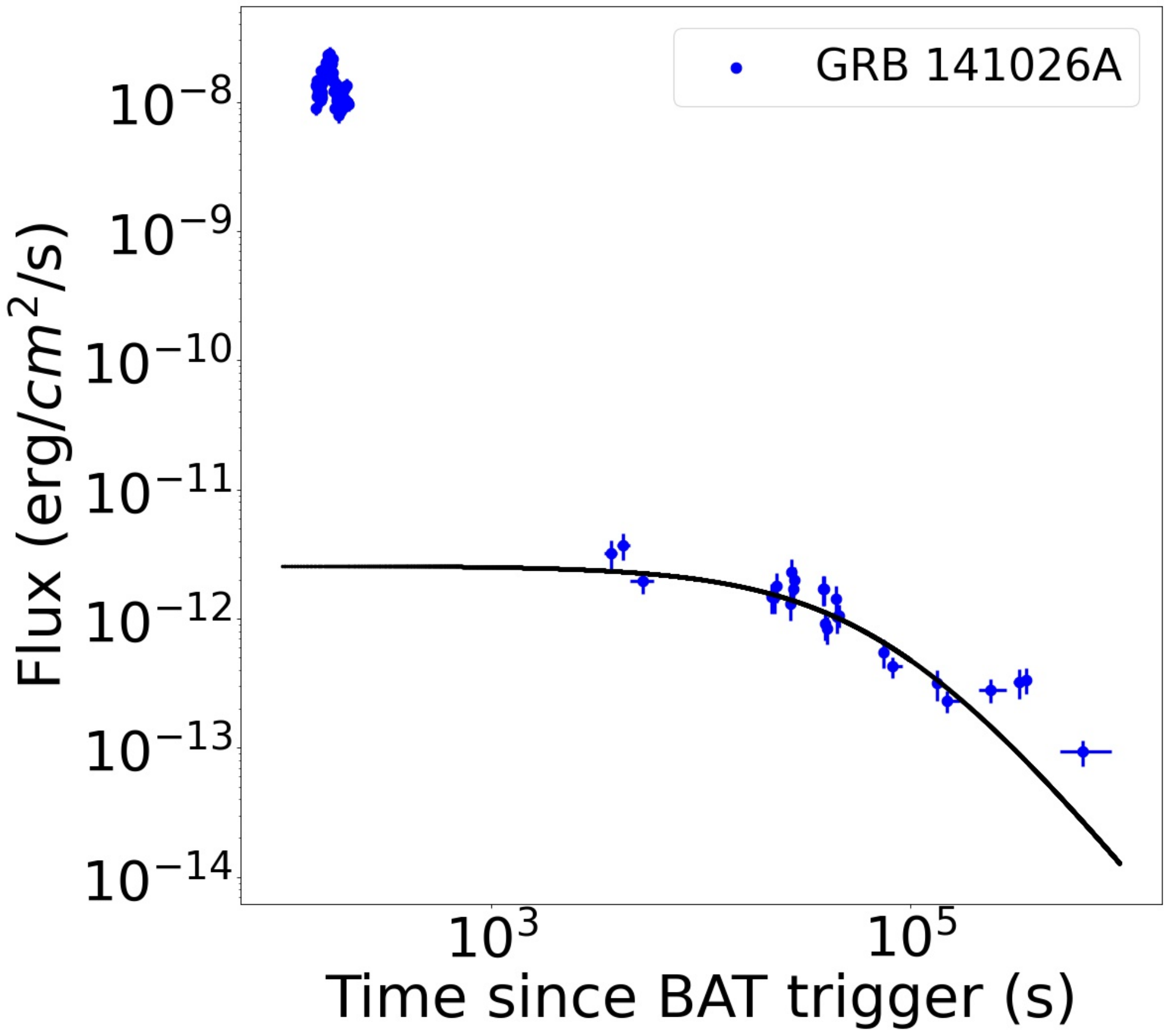}
	\includegraphics[width=0.31\hsize]{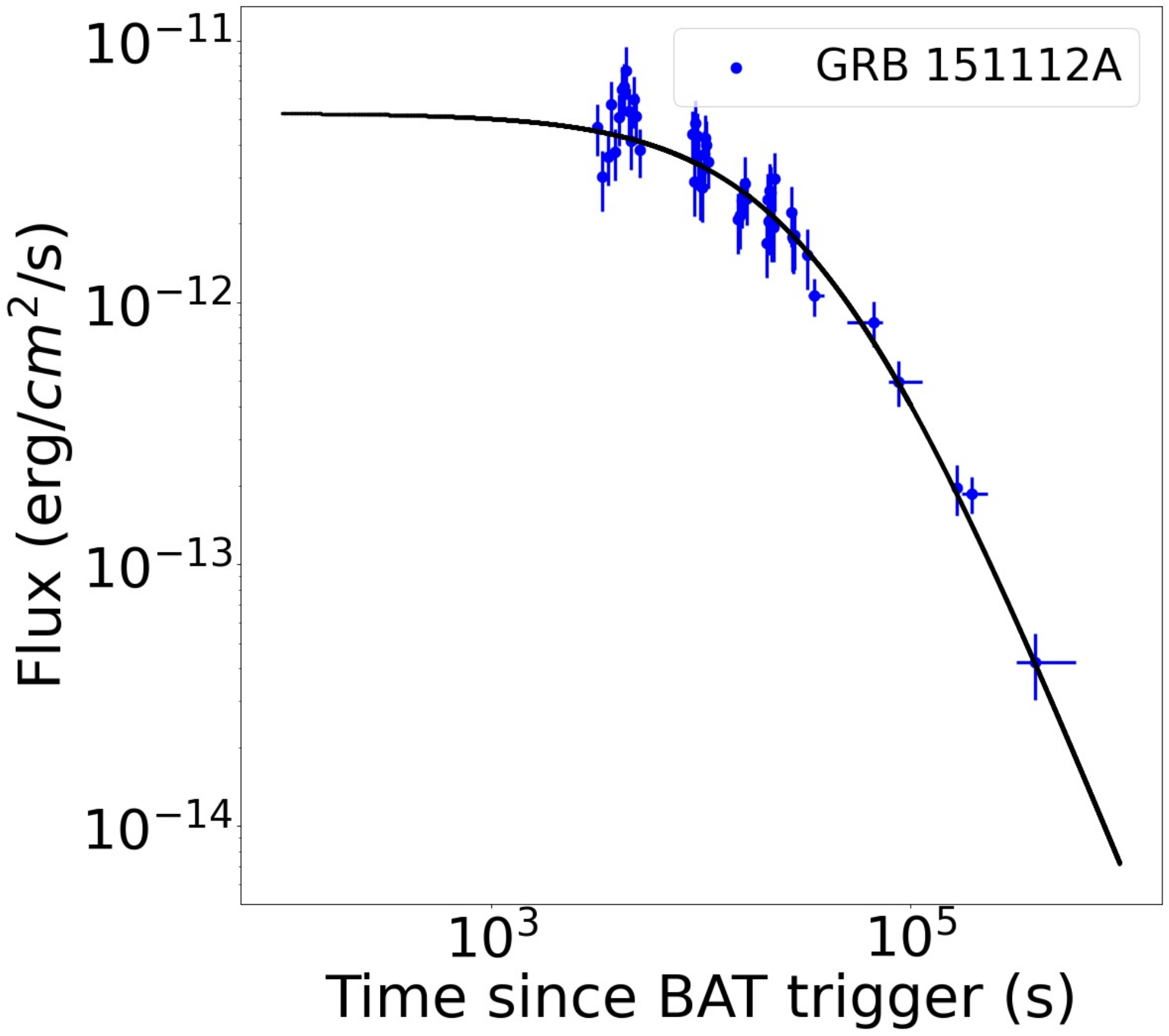}
	\includegraphics[width=0.31\hsize]{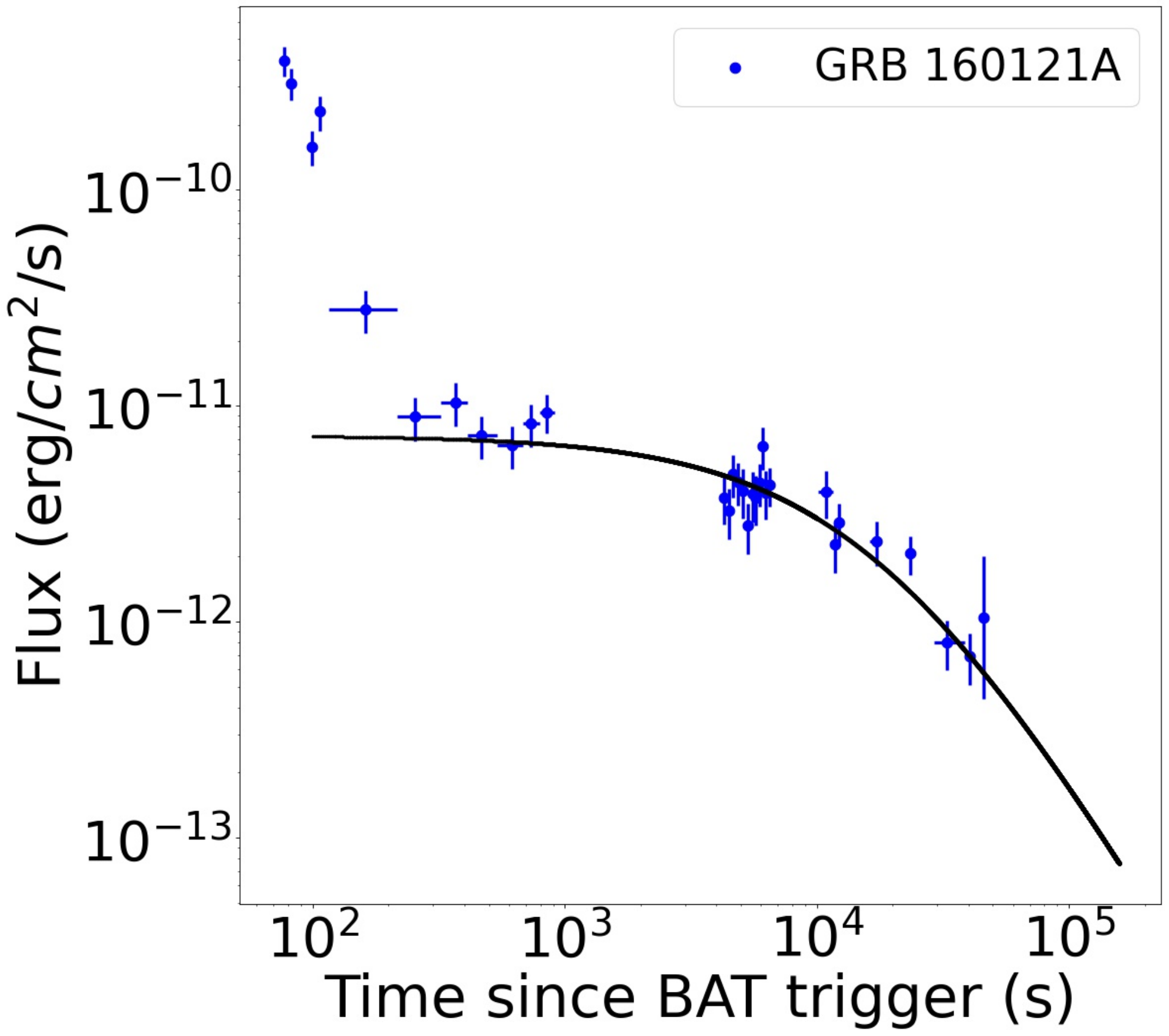}
	\includegraphics[width=0.31\hsize]{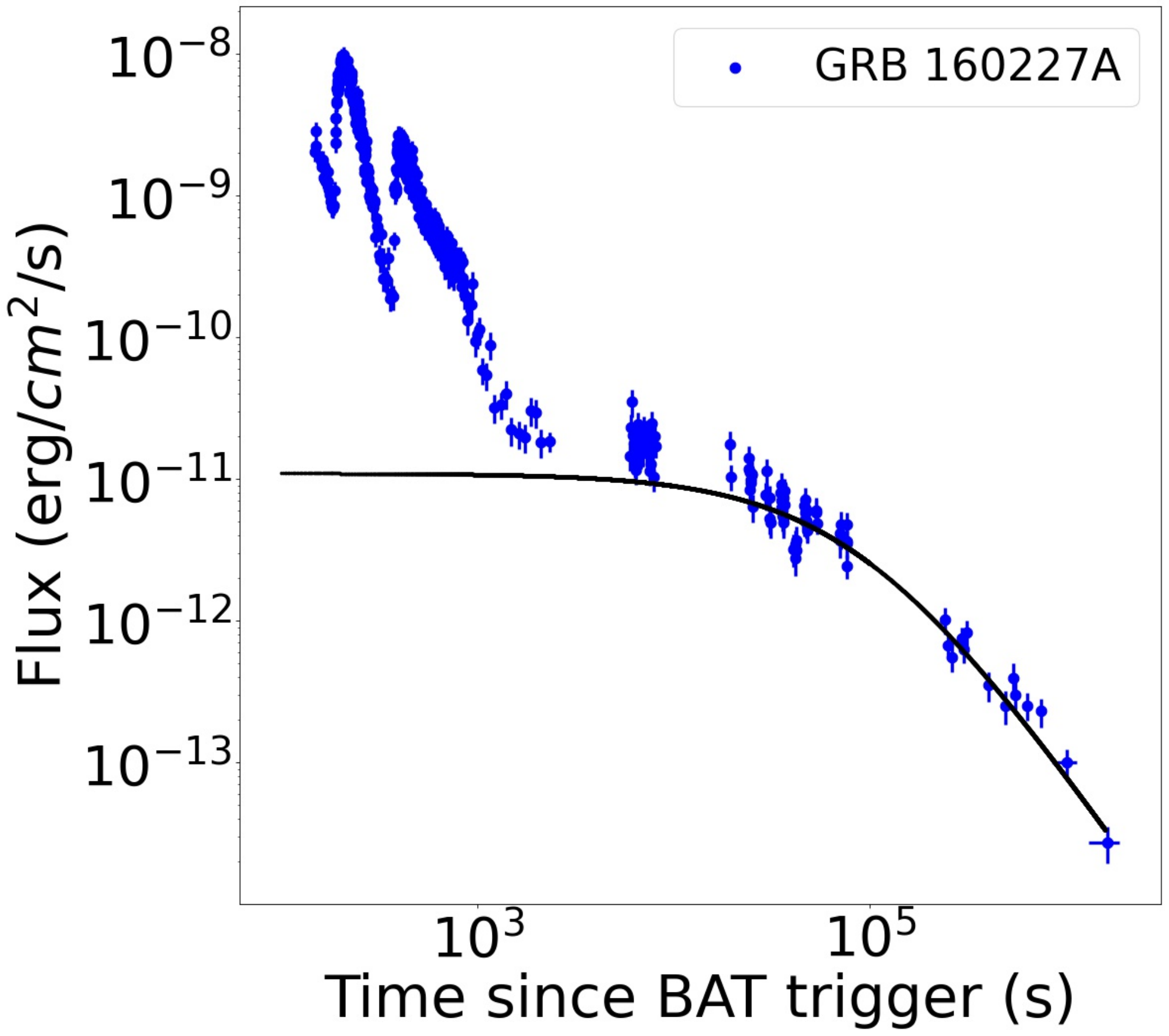}
	\includegraphics[width=0.31\hsize]{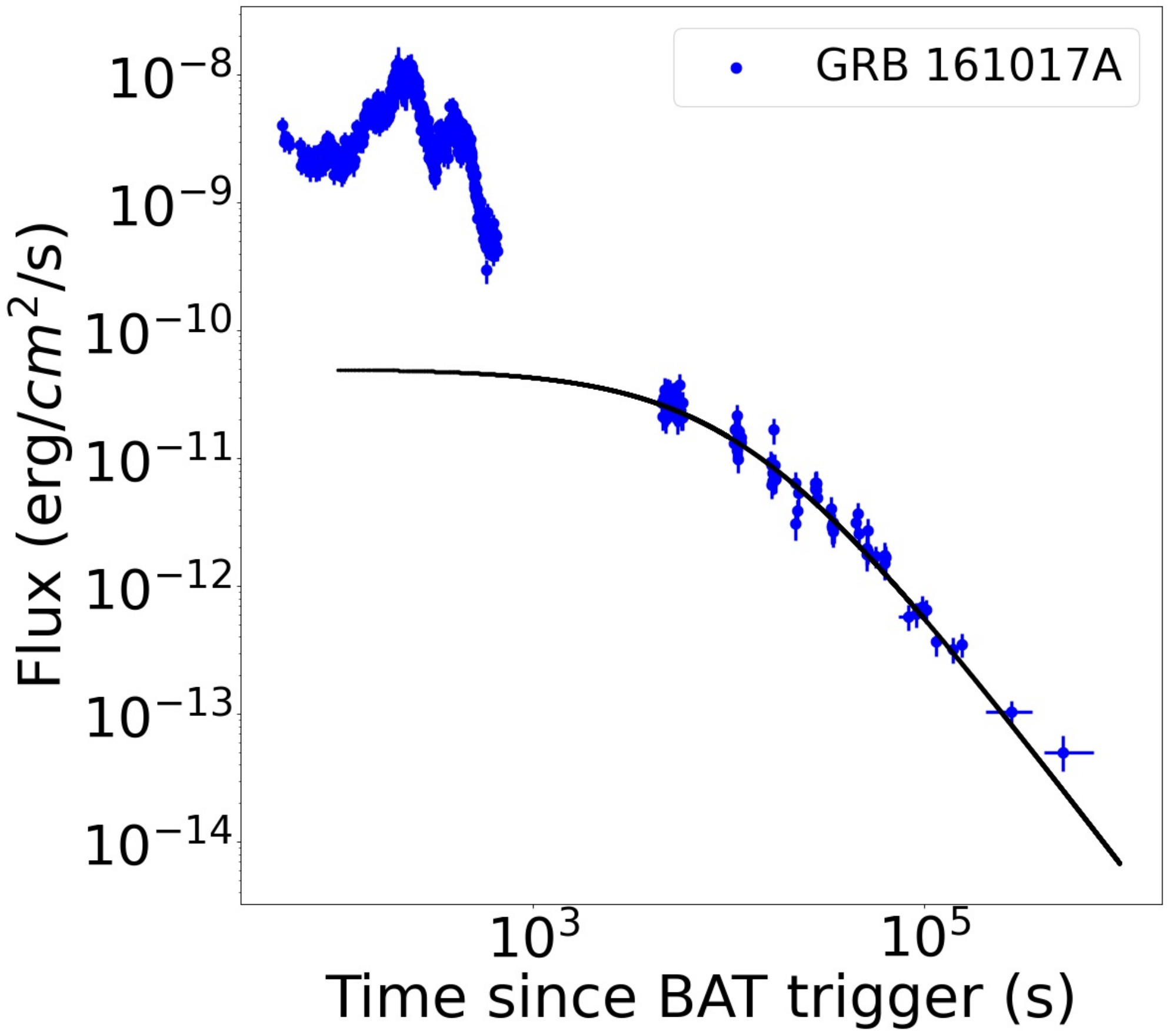}
	\includegraphics[width=0.31\hsize]{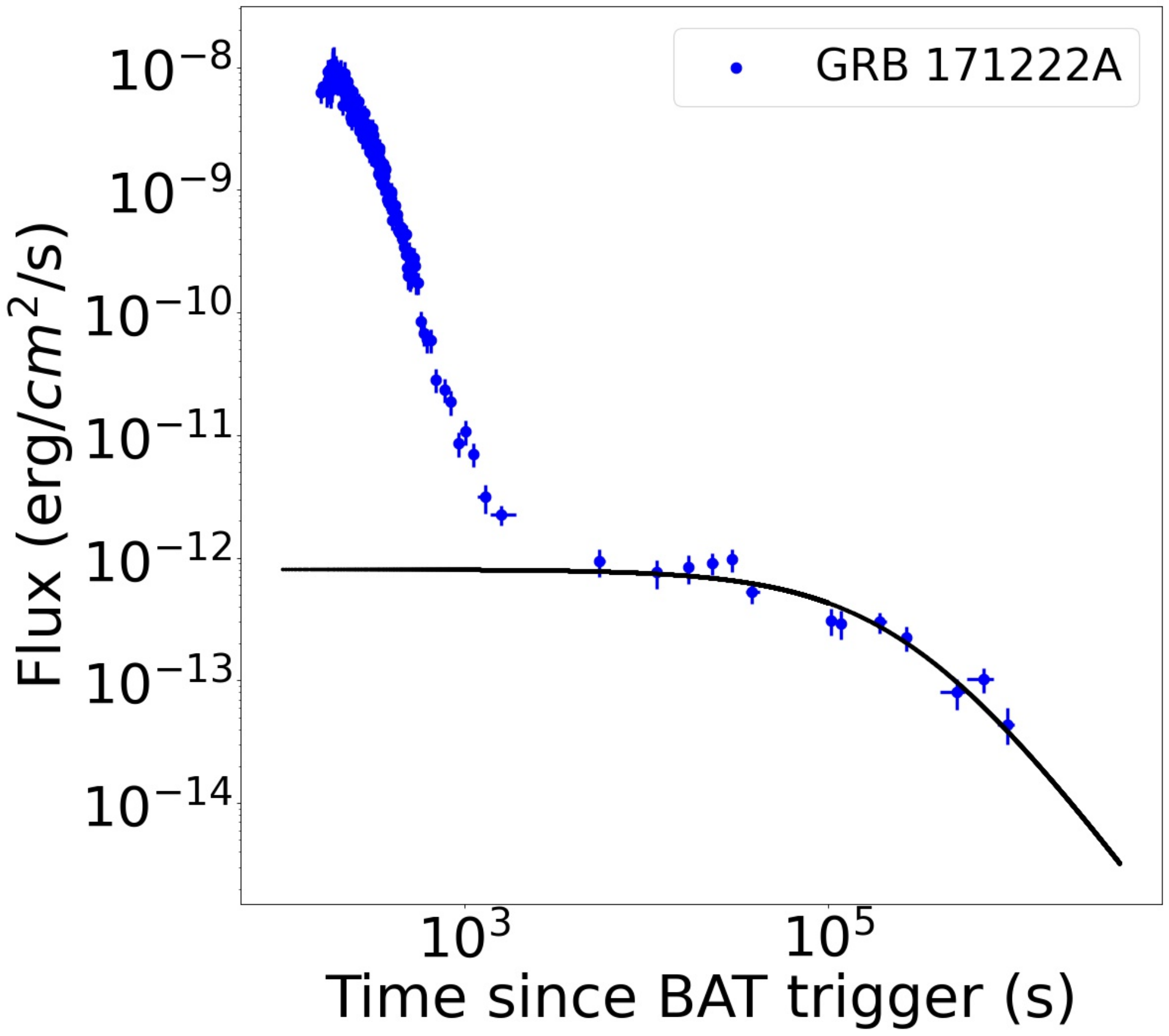}
	\includegraphics[width=0.31\hsize]{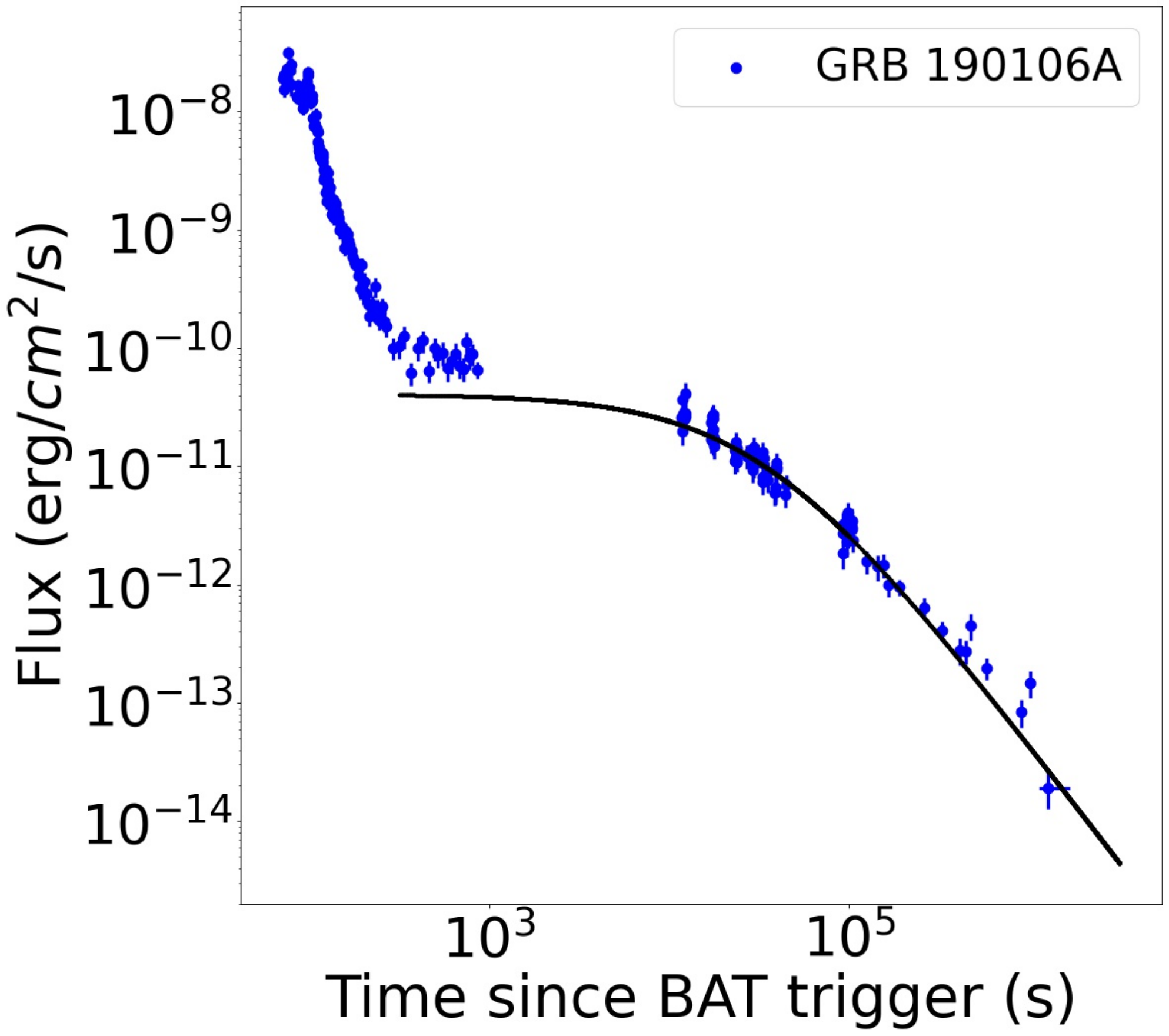}
	\includegraphics[width=0.31\hsize]{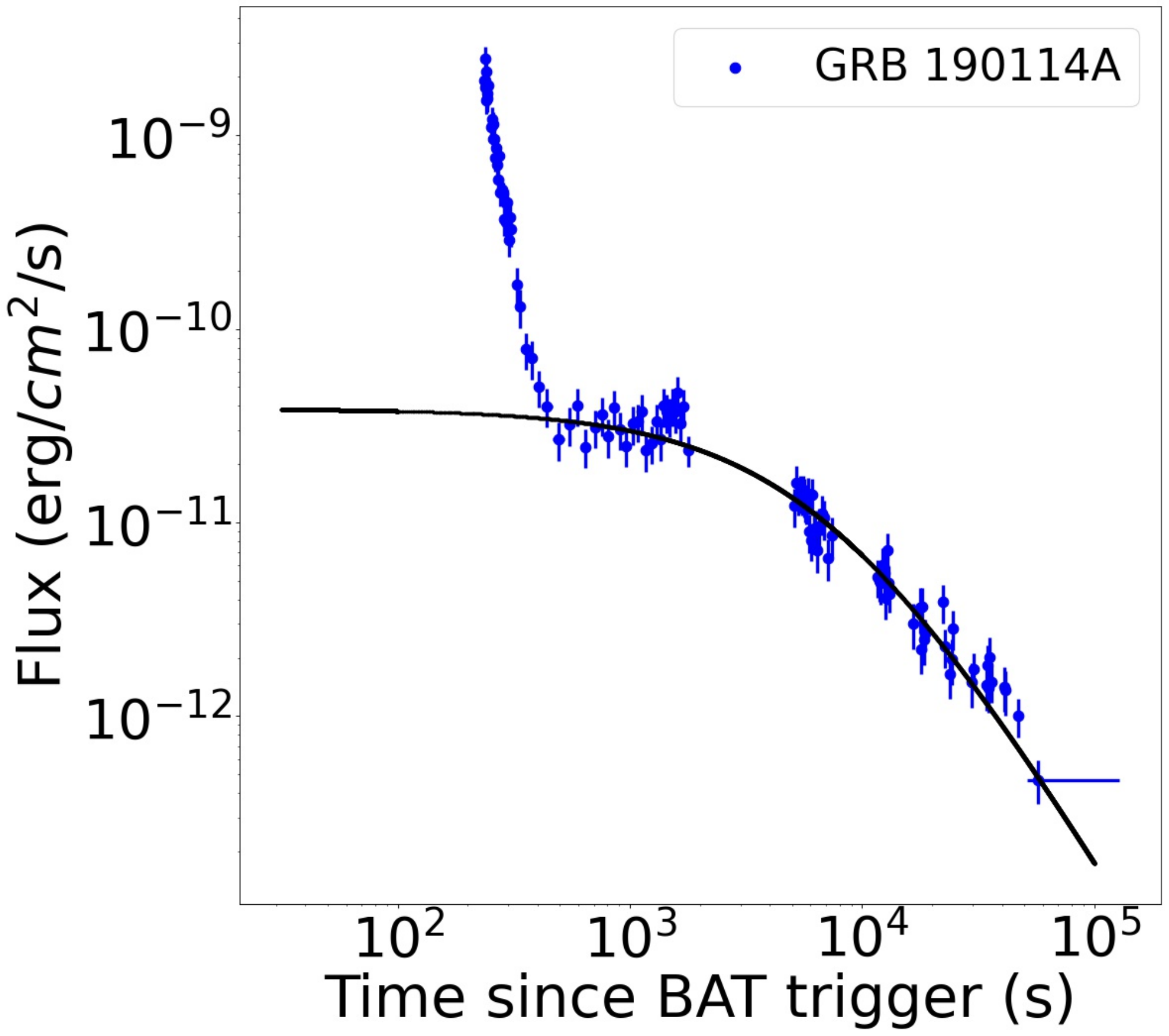}
	\includegraphics[width=0.31\hsize]{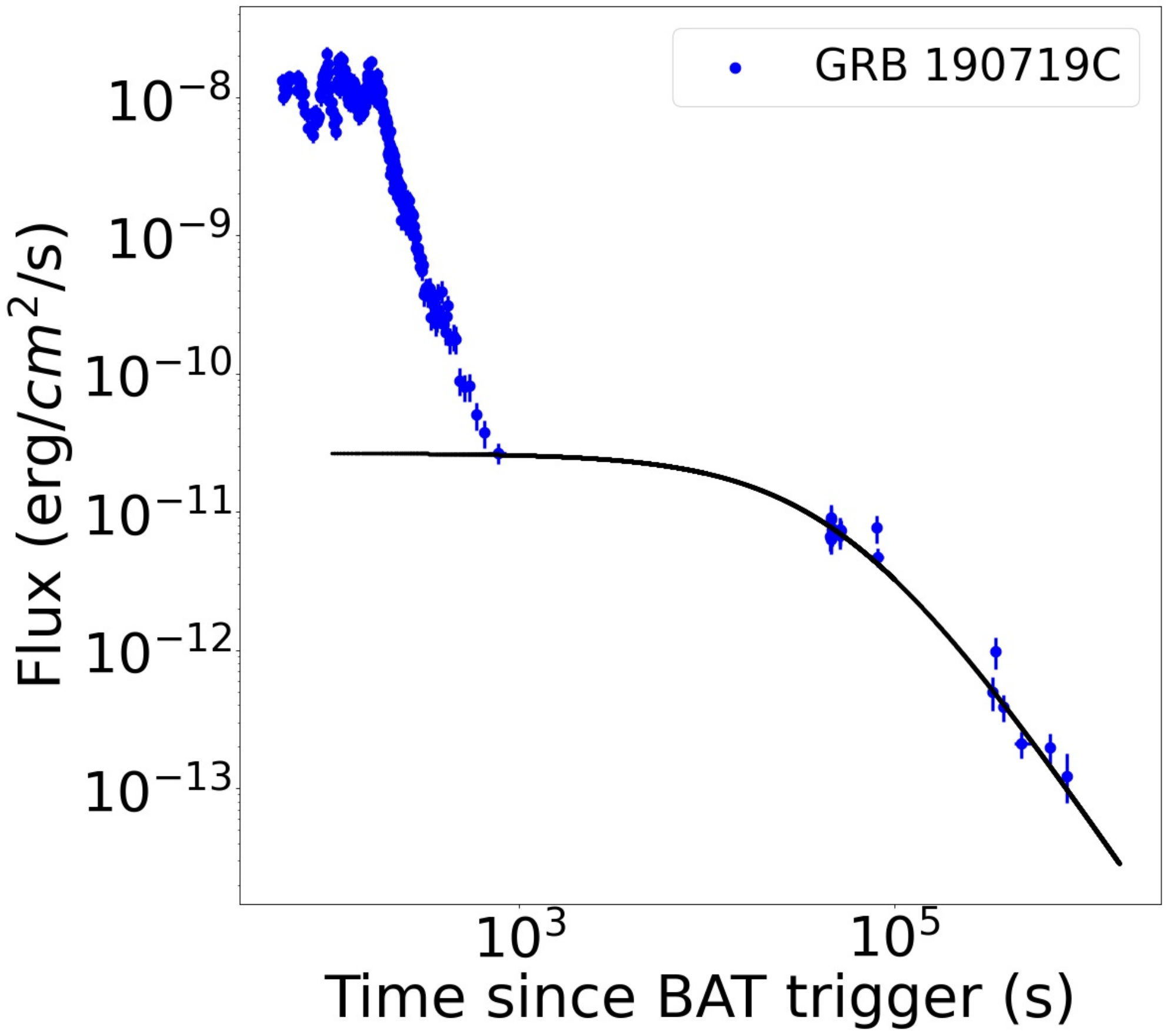}
	\caption{(Continued)}
	\label{silvers2}
\end{figure*}

\begin{figure}
	\centering
	\includegraphics[width=0.5\textwidth]{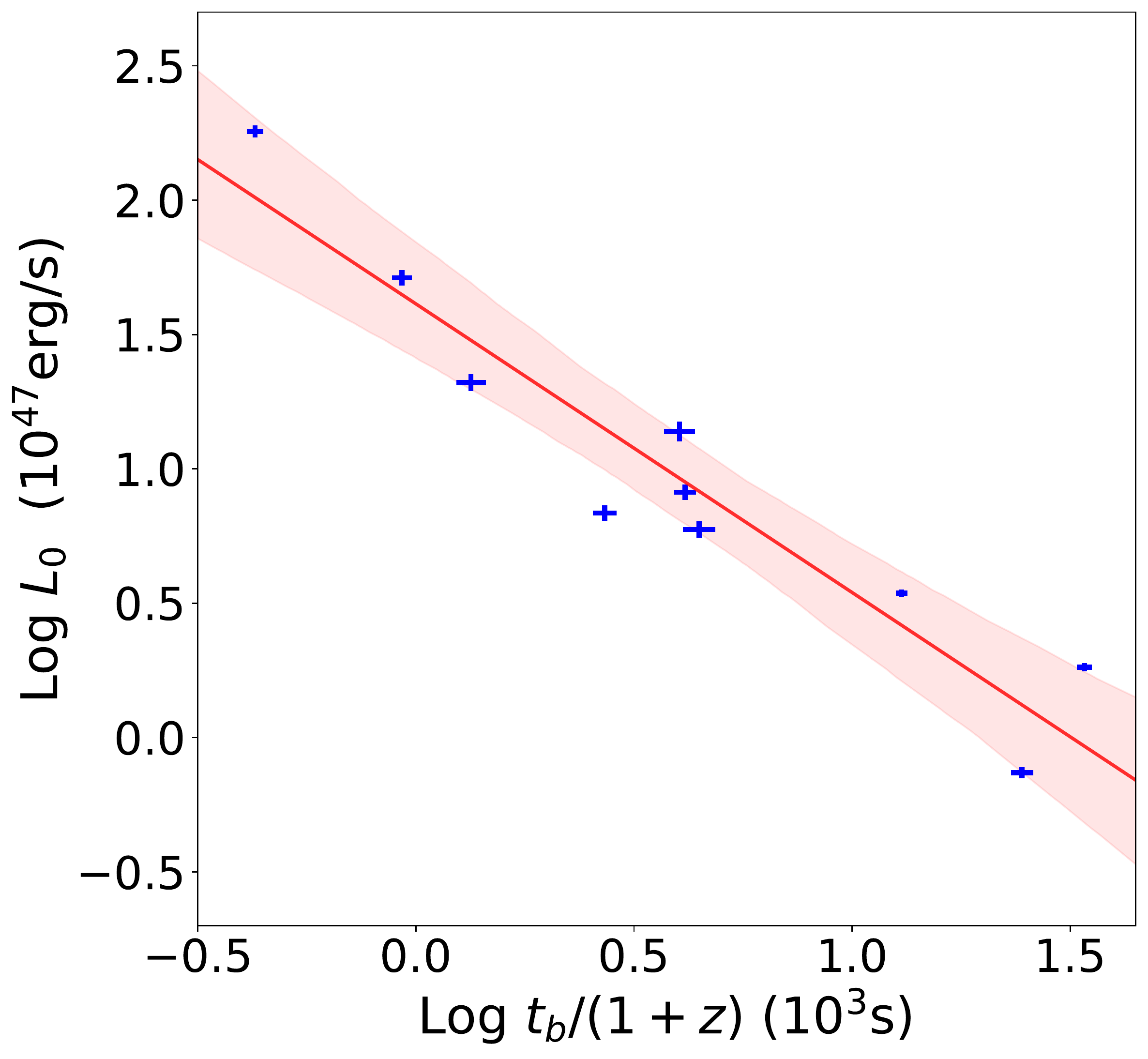}
	\caption{\label{Fig3} Correlation between luminosity $L_0$ and the end time $t_b$ of the plateau in X-ray light curves of GRBs.
		The luminosity is obtained from the measured flux assuming
		a flat $\Lambda$CDM model with $\Omega_{m}$ = 0.3 and
		$H_{0}$ = 70 km/s/Mpc. The term $(1+z)$ is the relativistic time dilation factor to transfer the time into the source's rest
		frame.
		Blue points are the long GRBs in Gold sample.
		The red line is the best fit with $k= 1.07_{-0.12}^{+0.13}$ and $b=1.61_{-0.11}^{+0.10}$ ($1\sigma$) with intrinsic scatter
		$\sigma_{\rm int}=0.22^{+0.08}_{-0.05}$. The shad region shows the $2\sigma$ confidence level. The plateau luminosity is inversely
		proportional to the timescale of the energy injection. Therefore, the energy reservoir should be almost a constant, which strongly
		supports the newly born magnetar can be treated as standard candle. }
\end{figure}

%%%%%%%%%%%%%%%%%%%%%%%%%%%%%%%%%%%%%%%%%%%%%%%%%%
%Fig.2
\begin{figure}
	\includegraphics[width=\textwidth]{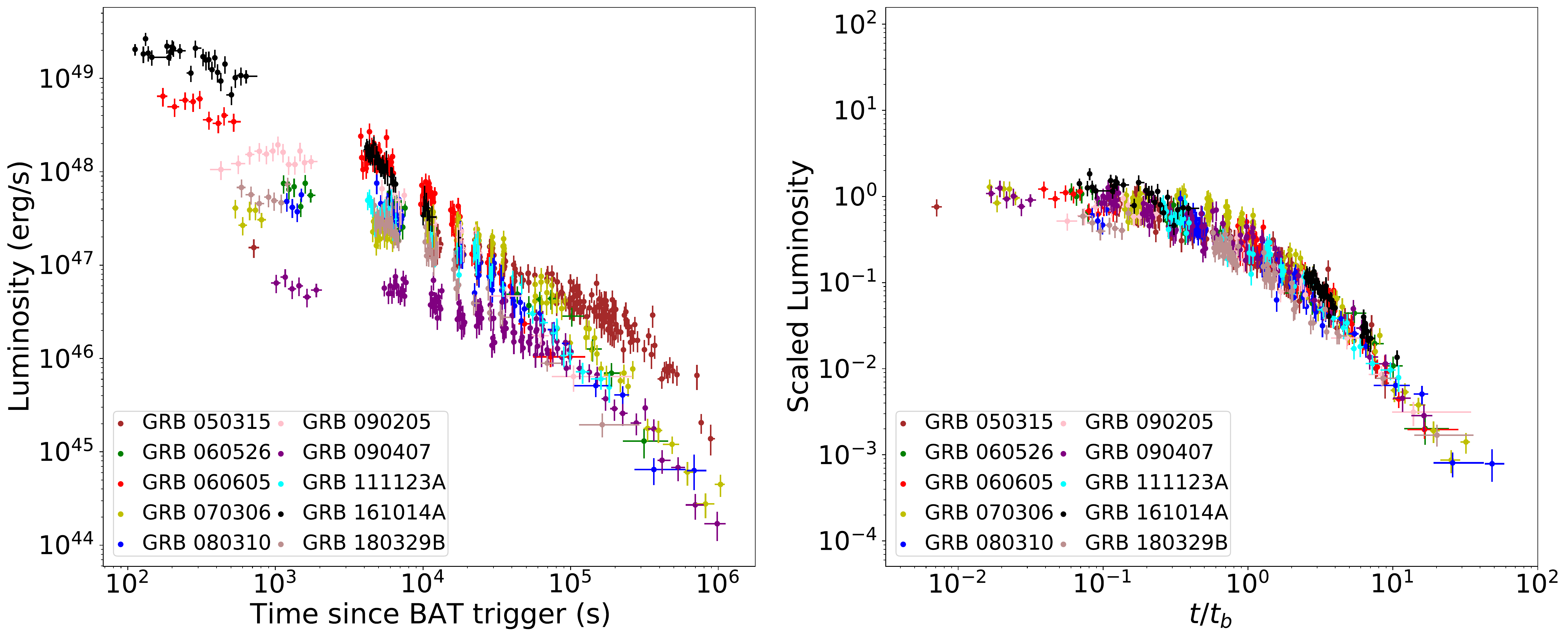}
	\caption{\label{Fig4} Original and scaled plateau light curves of GRBs in the Gold sample. Left panel shows the original X-ray
		(0.3-10 keV) light curves of long GRBs. Right panel shows the scaled light curves of the same GRBs. We first scale the spin-down time scale $t_b$ at the same point using $t/t_b$ for
		different GRBs, where $t$ is the horizon axis of left panel. From equation (\ref{eq:logL}), the corresponding value of $L_0$ can be obtained employing $t_b$. The scaled luminosity equals to the observed one divided by $L_0$. The scaled light curves show a universal behavior. The scale method is similar to that employed to standardize type Ia supernovae. The dispersion of the scaled luminosity is 0.5 dex.}
\end{figure}

%%%%%%%%%%%%%%%%%%%%%%%%%%%%%%%%%%%%%%%%%%%%%%%%%%
%Fig.3
\begin{figure}
	\includegraphics[width=0.5\textwidth]{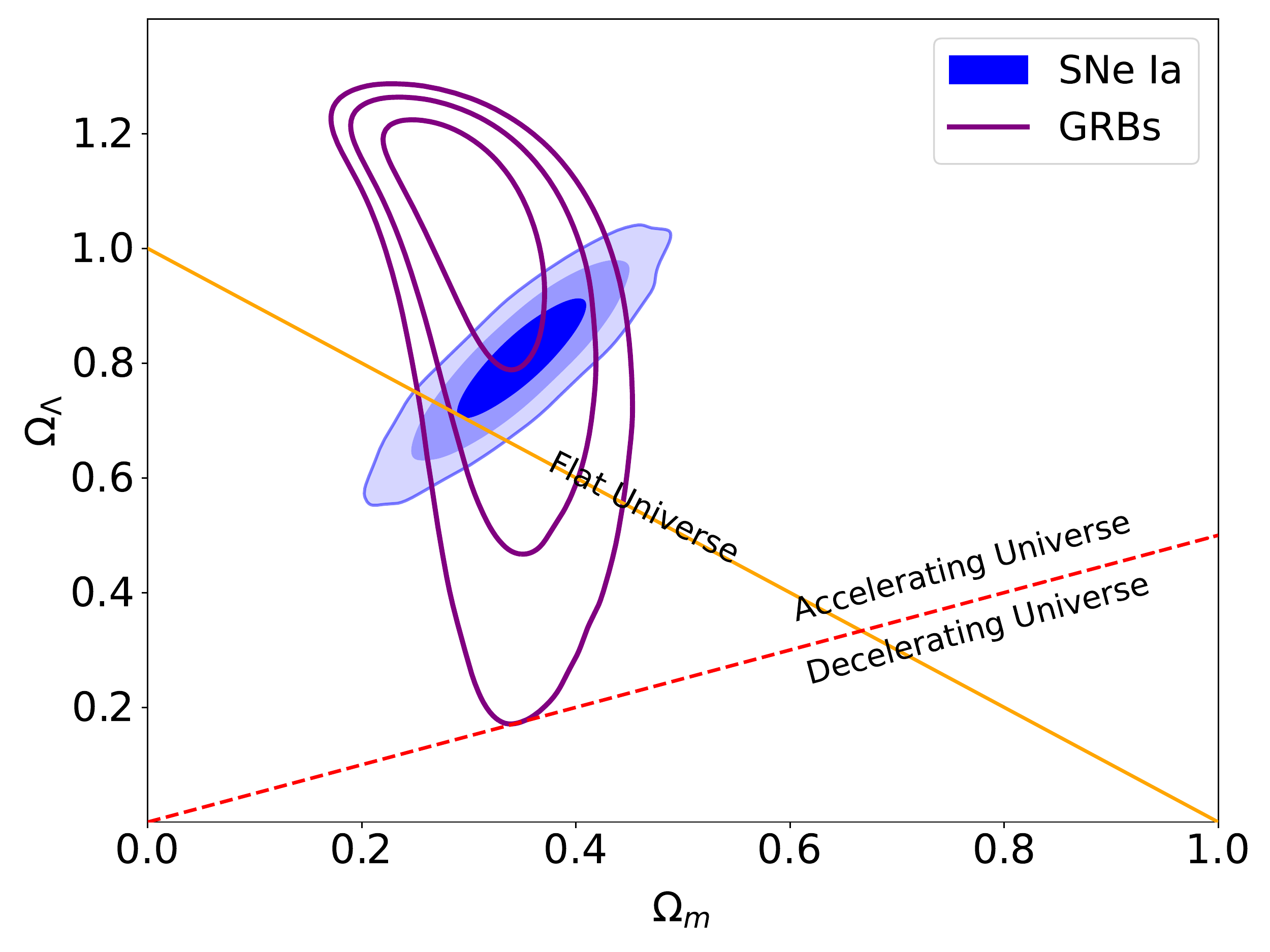}
	\includegraphics[width=0.56\textwidth]{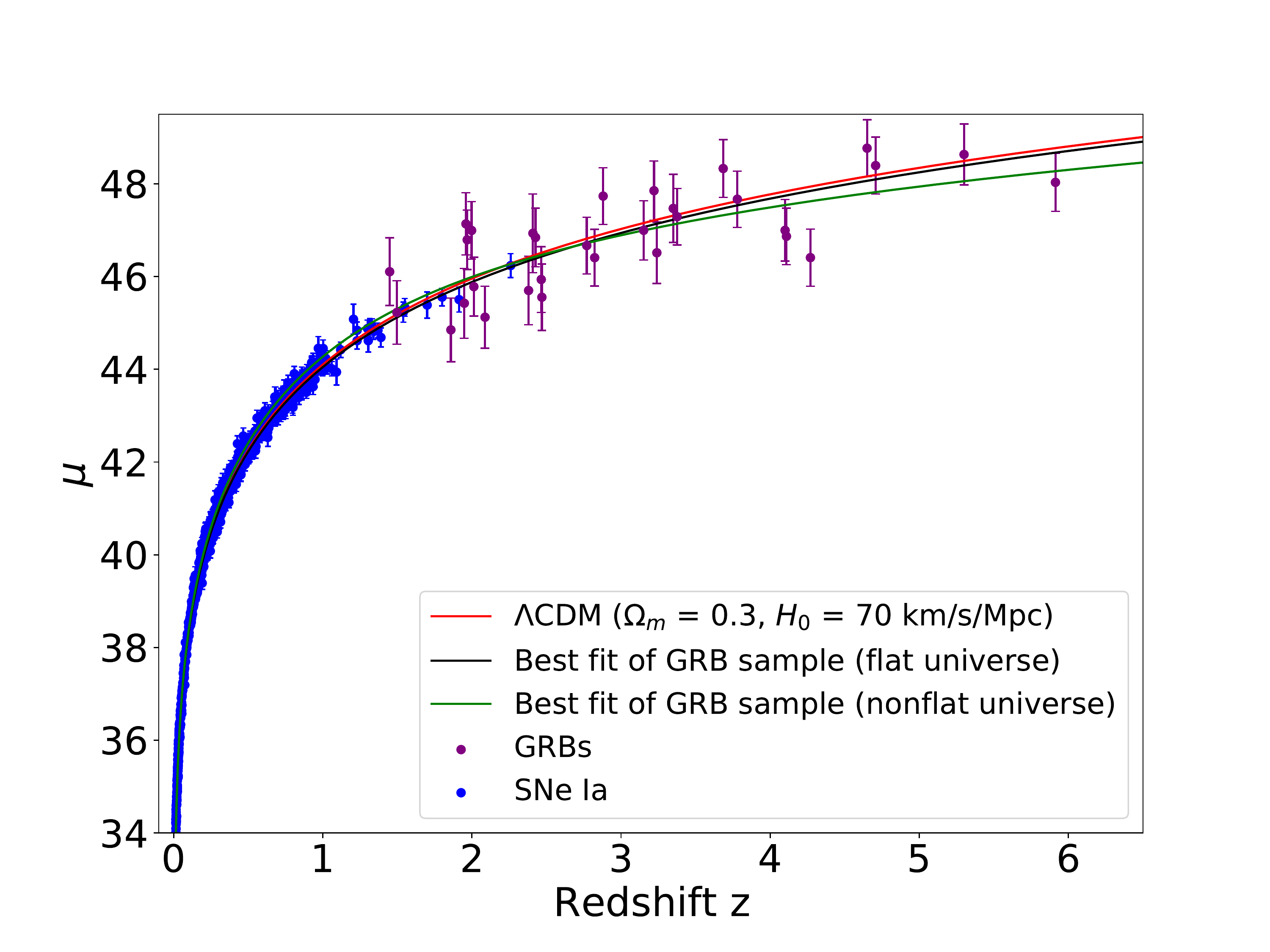}
	\caption{\label{Fig5} Evidence for dark energy from GRB constraint. Left panle shows the constraints ($1\sigma$ to $3\sigma$) on cosmic matter
		density $\Omega_m$ and dark energy density $\Omega_{\Lambda}$ in nonflat $\Lambda$CDM model using 31 calibrated GRBs in total sample (purple contours) and the Pantheon sample (blue contours). The best fits are $\Omega_m =0.32^{+0.05}_{-0.10}$ and $\Omega_{\Lambda}$ = $1.10^{+0.12}_{-0.31}$ (1$\sigma$) from GRBs in the poorly-explored high redshift range. The orange line indicates the flat universe $\Omega_m+\Omega_\Lambda=1$. The red dashed line shows the separation between accelerating and decelerating universe, i.e., the deceleration parameter $q_0$ is equal to
		0. The evidence of nonzero $\Omega_\Lambda$ from the GRB sample is $3\sigma$. Right panel: the Hubble diagram of GRBs and supernovae.
		Blue points are supernovae from the
		Pantheon sample. Purple points are 31 long GRBs with $1\sigma$ uncertainty. The red solid line is a flat $\Lambda$CDM model with
		$\Omega_m=0.3$ and Hubble constant $H_0=70 $ km s$^{-1}$ Mpc$^{-1}$. For flat $\Lambda$CDM model, the best fit from the calibrated GRB sample is shown as black line with $\Omega_m=0.34 \pm 0.05$.
		The green solid line shows the best fit from 31 GRBs for nonflat $\Lambda$CDM model.}
\end{figure}

%%%%%%%%%%%%%%%%%%%%%%%%%%%%%%%%%%%%%%%%%%%%%%%%%%
%Fig.4

\begin{figure}
	\includegraphics[width=0.5\textwidth]{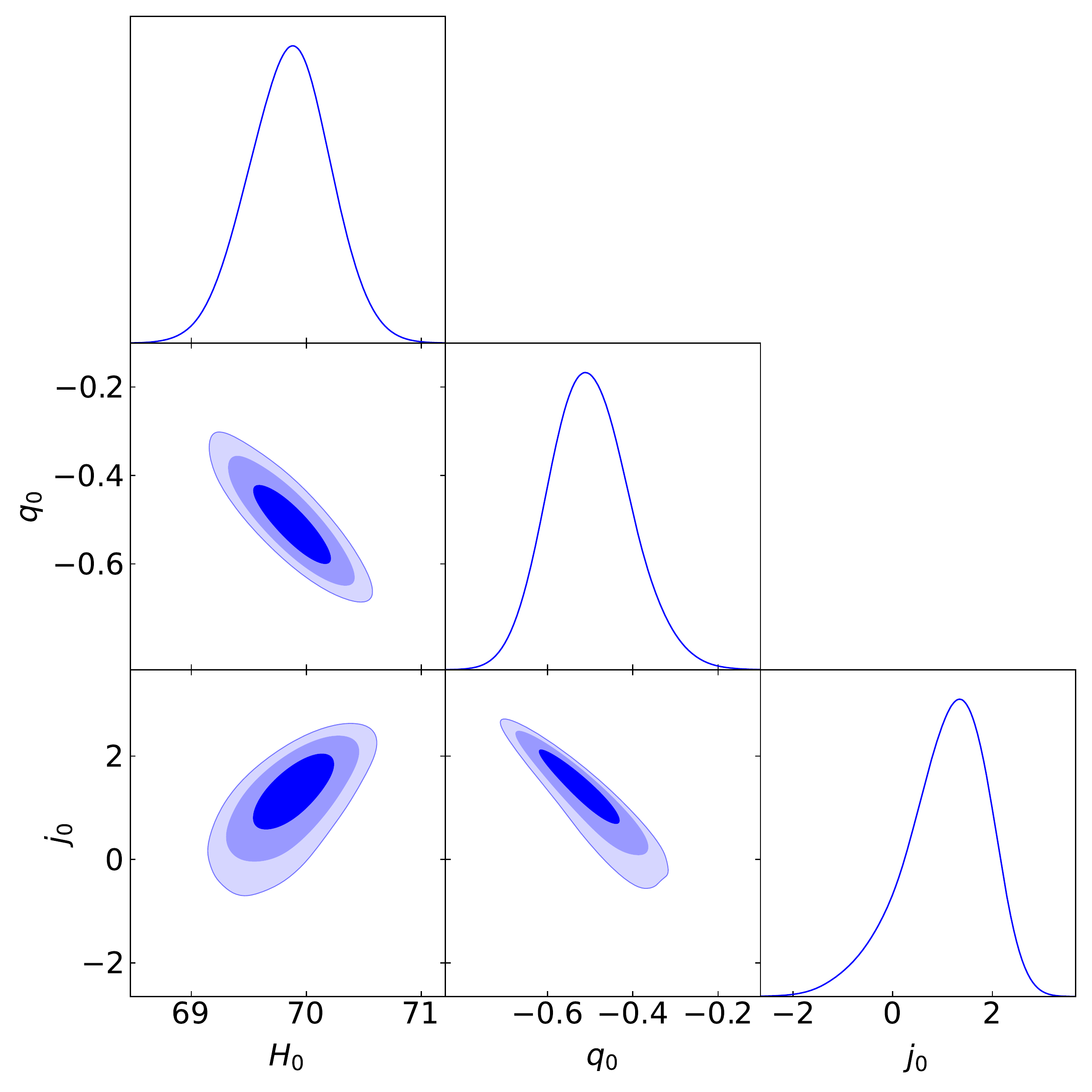}
	\includegraphics[width=0.5\textwidth]{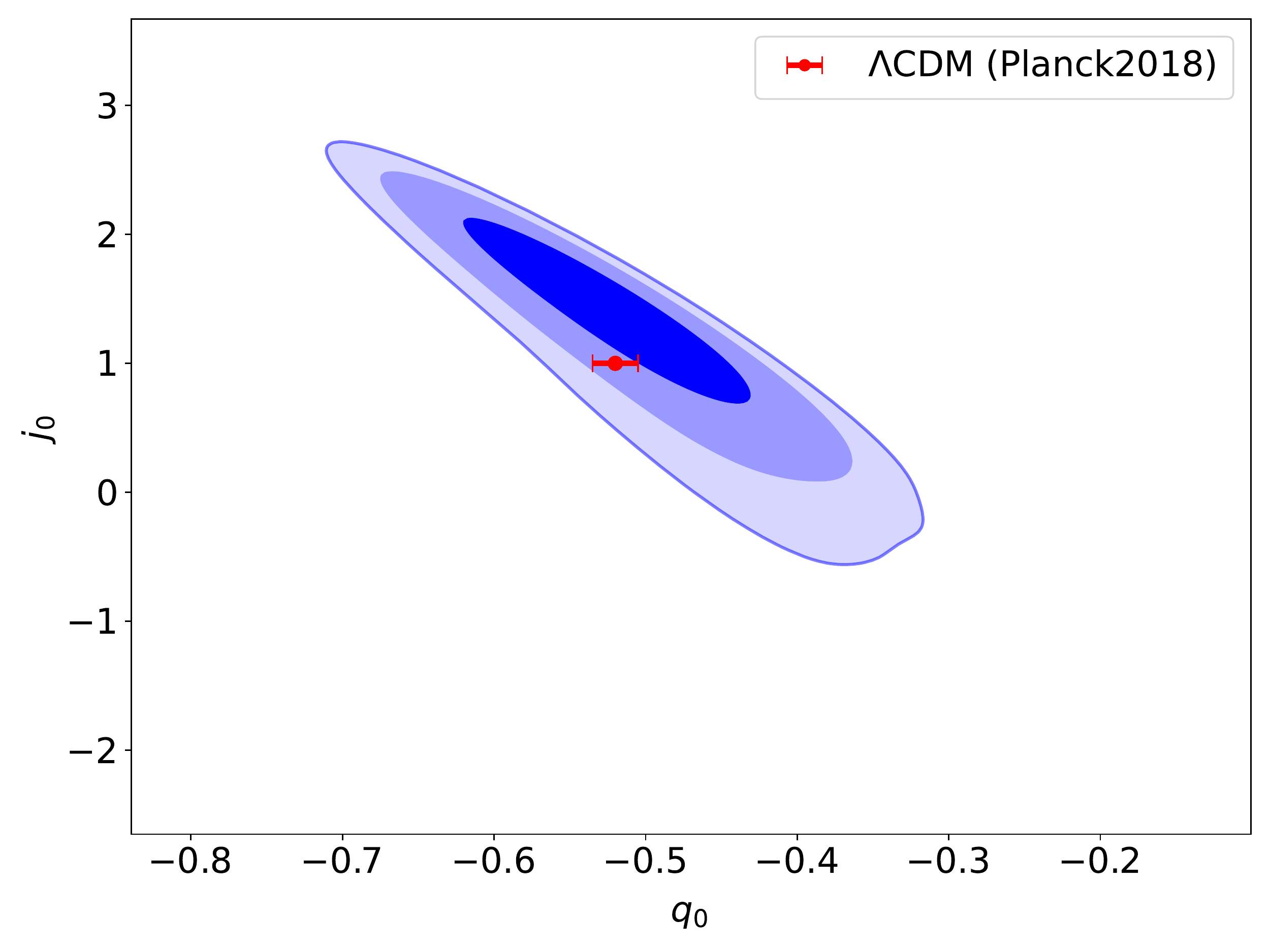}
	\caption{\label{Fig6} Constraints on the parameters $H_{0}$, $q_{0}$ and $j_{0}$ using 31 calibrated GRBs and supernovae in
		Pantheon sample. Left panel: confidence contours ($1\sigma,2\sigma$ and $3\sigma$) and marginalized likelihood distributions for
		$H_{0}$, $q_{0}$ and $j_{0}$. Right panel: Confidence regions in the $q_{0}$ and $j_{0}$ plane as provided calibrated GRBs and supernovae in
		Pantheon sample. Red point (-0.49, 1) represents the value given by Planck 2018 results \citep{2020A&A...641A...6P}. The values of $q_0$
		and $j_0$ are consistent with the predictions of  $\Lambda$CDM model at high redshifts in $2\sigma$ confidence level.}
\end{figure}

%%%%%%%%%%%%%%%%%%%%%%%%%%%%%%%%%%%%%%%%%%%%%%%%%%
%Tab.1

\linespread{1.0}
\begin{deluxetable*}{lccccclc}
	\small
	%\tablenum{1}
	\tablecaption{Fitting results and some key parameters of the MD-SGRBs and GW-LGRBs samples. \label{T1}}
	\tablewidth{0pt}
	\tablehead{
		\colhead{Name} & \colhead{$z$$^{a}$} & \colhead{$T_{90}$$^{b}$} & \colhead{$F_0$$^{c}$} &
		\colhead{$t_{b}$$^{c}$} & \colhead{$\chi^{2}_{r}$} & \colhead{$\gamma$$^{d}$} & \colhead{$\mu_{obs}$$^{e}$} \\
		\colhead{}&\colhead{}&\colhead{(s)}&\colhead{10$^{-11}$(erg/cm$^{2}$/s)} &\colhead{10$^{3}$(s)}&\colhead{}&\colhead{}&\colhead{}
	}
	\startdata
	\textbf{Gold} & & & & & & & \\
	\hline
	GRB 050315&1.95&95.6&0.76$_{-0.03}^{+0.03}$&100.61$_{-3.93 }^{+4.06 }$&1.07&1.89\tnote{(1)}&45.42$\pm$0.75\\
	GRB 060526&3.22&298.2&0.76$_{-0.05}^{+0.06}$&18.82$_{-1.56 }^{+1.63 }$&1.19&1.89\tnote{(1)}&47.85$\pm$0.64\\
	GRB 060605&3.78&79.1&4.54$_{-0.29}^{+0.31}$&4.44$_{-0.23 }^{+0.24 }$&1.39&1.89\tnote{(1)}&47.66$\pm$0.61\\
	GRB 070306&1.50&209.5&2.93$_{-0.09}^{+0.09}$&32.44$_{-1.00 }^{+1.02 }$&1.94&1.80\tnote{(2)}&45.22$\pm$0.68\\
	GRB 080310&2.43&365.0&1.58$_{-0.10}^{+0.10}$&14.19$_{-0.79 }^{+0.83 }$&1.11&2.09&46.84$\pm$0.63\\
	GRB 090205&4.65&8.8&0.85$_{-0.06}^{+0.06}$&7.56$_{-0.56 }^{+0.62 }$&1.07&2.07\tnote{(2)}&48.76$\pm$0.61\\
	GRB 090407&1.45&310.0&0.47$_{-0.02}^{+0.02}$&60.04$_{-3.36 }^{+3.65 }$&1.15&2.22\tnote{(2)}&46.10$\pm$0.73\\
	GRB 111123A&3.15&290.0&0.72$_{-0.06}^{+0.06}$&16.69$_{-1.31 }^{+1.40 }$&1.29&2.55&46.99$\pm$0.64\\
	GRB 161014A&2.82&18.3&33.81$_{-1.70}^{+1.72}$&1.63$_{-0.07 }^{+0.07 }$&0.89&1.83&46.40$\pm$0.61\\
	GRB 180329B&2.00&210.0&2.75$_{-0.18}^{+0.19}$&8.12$_{-0.50 }^{+0.54 }$&0.84&1.87& 46.99$\pm$0.62\\
	\hline \hline
	\textbf{Silvers}& & & & & &  &\\
	\hline
	GRB 050319&3.24&152.5&1.55$_{-0.08}^{+0.08}$&33.25$_{-2.18 }^{+2.29 }$&0.90&1.85\tnote{(1)}&46.51$\pm$0.66\\
	GRB 050505&4.27&58.9&3.61$_{-0.18}^{+0.19}$&13.89$_{-0.57 }^{+0.60 }$&0.84&2.09\tnote{(1)}&46.40$\pm$0.62\\
	GRB 050814&5.30&150.9&0.34$_{-0.04}^{+0.04}$&37.27$_{-1.99 }^{+2.15 }$&1.19&1.97&48.63$\pm$0.66\\
	GRB 051008&2.77&$>$32.0&5.86$_{-0.41}^{+0.45}$&6.03$_{-0.33 }^{+0.33 }$&1.36&1.95&46.66$\pm$0.61\\
	GRB 060906&3.69&43.5&0.59$_{-0.04}^{+0.04}$&12.75$_{-0.92 }^{+1.00 }$&1.74&2.10\tnote{(2)}&48.33$\pm$0.62\\
	GRB 061222A&2.09&71.4&4.33$_{-0.19}^{+0.21}$&29.93$_{-1.08 }^{+1.11 }$&1.06&1.84\tnote{(2)}&45.12$\pm$0.67\\
	GRB 081008&1.97&185.5&3.23$_{-0.42}^{+0.50}$&7.28$_{-0.79 }^{+0.83 }$&1.10&1.98&46.79$\pm$0.64\\
	GRB 090516&4.11&140.0&3.88$_{-0.22}^{+0.25}$&9.17$_{-0.45 }^{+0.43 }$&1.16&2.03&46.86$\pm$0.61\\
	GRB 100424A&2.47&104.0&523.12$_{-28.08}^{+30.55}$&0.19$_{-0.01 }^{+0.01 }$&1.57&1.66&45.93$\pm$0.71\\
	GRB 120404A&2.88&38.7&4.58$_{-0.37}^{+0.39}$&3.26$_{-0.22 }^{+0.24 }$&1.03&1.90&47.73$\pm$0.61\\
	GRB 130606A&5.91&276.6&2.76$_{-0.31}^{+0.36}$&8.23$_{-0.77 }^{+0.79 }$&0.90&1.86&48.03$\pm$0.62\\
	GRB 140518A&4.71&60.5&2.75$_{-0.19}^{+0.20}$&3.26$_{-0.23 }^{+0.25 }$&1.33&2.09\tnote{(2)}&48.39$\pm$0.61\\
	GRB 141026A&3.35&146.0&0.26$_{-0.03}^{+0.04}$&75.72$_{-10.62 }^{+11.79 }$&\textbf{1.19}&1.92&47.47$\pm$0.73\\
	GRB 151112A&4.10&19.3&0.53$_{-0.03}^{+0.03}$&38.35$_{-3.23 }^{+3.42 }$&0.91&2.28\tnote{(2)}&46.99$\pm$0.66\\
	GRB 160121A&1.96&12.0&0.73$_{-0.06}^{+0.07}$&18.03$_{-2.52 }^{+2.95 }$&0.98&2.21&47.13$\pm$0.67\\
	GRB 160227A&2.38&316.5&1.10$_{-0.05}^{+0.07}$&92.41$_{-6.16 }^{+4.94 }$&1.84&1.67\tnote{(2)}& 45.70$\pm$0.74\\
	GRB 161017A&2.01&216.3&5.04$_{-0.35}^{+0.39}$&11.73$_{-0.71 }^{+0.75 }$&1.07&1.99& 45.78$\pm$0.63\\
	GRB 171222A&2.41&174.8&0.08$_{-0.01}^{+0.01}$&268.70$_{-39.91 }^{+48.54 }$&1.24&1.99& 46.93$\pm$0.85\\
	GRB 190106A&1.86&76.8&4.10$_{-0.31}^{+0.36}$&33.13$_{-2.36 }^{+2.44 }$&1.03&1.95& 44.85$\pm$0.69\\
	GRB 190114A&3.38&66.6&3.87$_{-0.20}^{+0.21}$&7.19$_{-0.37 }^{+0.40 }$&1.01&1.83&  47.29$\pm$0.61\\
	GRB 190719C&2.47&185.7&2.66$_{-0.32}^{+0.35}$&53.71$_{-5.44 }^{+6.19 }$&0.82&1.54&45.55$\pm$0.72\\
	\enddata
	\tablecomments{
		\begin{itemize}	
			\footnotesize
			\item[] \textbf{Reference.} (1) \citep{2019ApJ...883...97Z}, (2) \citep{2007ApJ...662.1093W}.
			\item[a] The measured redshifts are adopted from the published papers and GCNs.
			\item[b] The duration of the GRBs are obtained from the Swift GRB table at https://swift.gsfc.nasa.gov/archive/grb$_{-}$table.html/
			\item[c] Physical parameters $F_0$ and $t_b$ are derived from equation (\ref{Lt}). The corresponding 1 $\sigma$ errors, $\sigma_{F_{0}}$ and $\sigma_{t_{b}}$, are derived from function $\sqrt{({\sigma_{u}}^2+{\sigma_{d}}^2)/2}$. Here, $\sigma_{u}$ and $\sigma_{d}$ are the upper error and the lower error of the parameters, respectively.
			\item[d] $\gamma=\beta+1$ is the photon index of the plateau phase. When $\gamma$ is not given in published papers, the value from the Swift GRB table is used.
			\item[e]  $\mu_{obs}$ is distance modulus derived from Equation (\ref{eq:muobs}) by using the calibrated Dainotti relation.
	\end{itemize}}
\end{deluxetable*}

\end{document}